%% file: manuscript.tex
\documentclass[10pt]{article}
\usepackage{geometry}
\usepackage{graphicx}
\usepackage[FIGTOPCAP,nooneline]{subfigure}
\usepackage[T1]{fontenc}
\usepackage{lmodern}
\usepackage{bm}
\usepackage{amsmath, amsfonts, amssymb}
\usepackage{natbib}
\newcommand{\citey}{\citeyearpar}
\usepackage{multirow}
\usepackage{afterpage, array, rotating}
\usepackage{url}
\usepackage{authblk}
\usepackage{color}
\usepackage{algorithm}
\usepackage[noend]{algorithmic}
\usepackage{bm}
\usepackage{mathtools}
\usepackage{ulem}
\usepackage{booktabs}
\usepackage{tabularx}
\usepackage{multirow}
\usepackage{rotating}

\newcolumntype{A}{>{\centering\arraybackslash}p{5cm}}
\newcolumntype{B}{>{\centering\arraybackslash}p{1.2cm}}
\newcolumntype{C}{>{\centering\arraybackslash}p{3cm}}
\newcolumntype{D}{>{\raggedright\arraybackslash}p{11cm}}
\newcolumntype{E}{>{\raggedleft\arraybackslash}p{1.2cm}}

\newcolumntype{X}{>{\raggedright\arraybackslash}p{13cm}}

\title{
Declining Modularity of Intellectual Bases During the Emergence of Research Areas
}

\author[1]{Kazuki Nakajima}
\author[2]{Yuya Sasaki}
\author[1]{Masaki Aida}

\affil[1]{Tokyo Metropolitan University}
\affil[2]{The University of Osaka}

\begin{document}
\date{}
\maketitle

\begin{abstract}
Understanding how research areas emerge can help identify nascent areas early and inform research strategy, yet how the intellectual base of a field restructures as an area takes shape remains unclear.
We hypothesize that the emergence of a research area is accompanied by the integration of largely separate knowledge communities, observable as a decline in the modularity of its co-citation network, which represents its intellectual base.
We propose a framework that tracks this modularity over time, evaluates the statistical robustness of its changes, and identifies the papers highly associated with the decline.
We applied it to three areas with different modes of growth: higher-order network science, superstring theory, and graph representation learning.
In all three, modularity declined in correspondence with each area's emergence or transformation, and in superstring theory, the decline aligns with an independently documented transition.
Further analysis of higher-order network science shows that its decline reflects a cross-disciplinary integration.
In graph representation learning, the gradual decline is followed by a rise, which we interpret as a re-differentiation after the emergence period.
Our results suggest that a decline in the modularity of a co-citation network can serve as a structural signature that retrospectively characterizes this integrative mode of emergence.
\end{abstract}


\input{main}

\input{main.bbl}
\newpage

\input{sm}

\renewcommand{\refname}{Supplementary References}

\input{sm.bbl}
\end{document}

%% file: main.tex
\section{Introduction}

Science and technology often advance discontinuously through the emergence of research areas, in which the boundaries of existing disciplines intersect and new intellectual structures form~\citep{baaden2024}.
The introduction of an innovative concept, the development of a new technology, or the publication of a landmark paper or review can prompt research communities that had developed largely separately to influence one another and eventually become integrated into a single research area~\citep{bettencourt2009, herrera2010}.
For example, behavioral economics, which brought insights from psychology into economics~\citep{kahneman2013}; network medicine, which approaches disease from the perspective of network science~\citep{barabasi2011}; and cognitive science, which spans psychology, linguistics, neuroscience, and computer science in the study of the mind~\citep{miller2003}, are all research areas that arose from the integration of concepts and methods from different disciplines.
Quantitatively understanding when and how such new research areas emerge and through what structural changes they mature can provide an empirical basis for identifying nascent areas early and for planning research strategies~\citep{fortunato2018, zeng2017, clauset2017}.

In bibliometrics, network analysis based on citations among papers has served as a basic method for characterizing the knowledge structure of a field.
A co-citation network, in which two papers are linked when a third paper cites both, has long been used to reflect the intellectual base of a field at a given time~\citep{small1973, small1974}. From its community structure, this approach has developed into science mapping, which identifies research topics and specialties~\citep{boyack2010, rosvall2010, trujillo2018}.
A body of work has also accumulated on detecting research areas and topics that are newly emerging in citation networks.
Representative approaches identify research fronts from citation and co-citation density and from topological measures~\citep{shibata2008, shibata2009, chen2004, chen2006}, detect bursts in the frequency of terms or topics~\citep{kleinberg2003, mane2004}, and identify emerging topics from the formation of co-citation clusters and surges in citation~\citep{small2014}; related work characterizes the formation and maturation of research areas as structural transitions in collaboration or knowledge networks~\citep{bettencourt2009, herrera2010, sun2013, salatino2017}.
As an attempt to quantify the degree of integration of the intellectual base itself, an indicator has been proposed that captures the interdisciplinarity of a field as a combination of citation diversity and network coherence~\citep{rafols2010}, although this characterizes the set of papers at a single point in time.
In contrast, Shwed and Bearman~\citey{shwed2010} treated the strength of the community structure as a quantity to be tracked: they followed the modularity of citation networks over time and interpreted its decline as the formation of scientific consensus on contested propositions.

How the strength of the community structure of the intellectual base changes as a research area emerges, however, remains unclear.
We hypothesize that the emergence of a new research area is accompanied by the integration of knowledge communities that were previously largely separate, observed as a decline in the modularity of the co-citation network.
In the early or reorganizing phases of an area, papers that draw on knowledge across the boundaries of existing communities appear and promote the integration of the intellectual base, that is, they lower its modularity.
We propose a framework to test whether this decline is in fact observed in correspondence with the emergence and transformation of research areas.
The framework tracks the temporal change of modularity together with its statistical uncertainty and quantifies the extent to which the citation behavior of individual papers corresponds to the decline.

To test the framework, we selected three research areas that differ in era, discipline, and growth pattern: higher-order network science, currently taking shape~\citep{battiston2020}; superstring theory, which underwent a known historical transition, the first superstring revolution, in the mid-1980s~\citep{green20121, green20122}; and graph representation learning, which grew rapidly in the late 2010s as the confluence of several methodological lineages~\citep{hamilton2020}.
We designed this contrast to examine whether the framework works across qualitatively different modes of emergence.
Across all three areas, we detected a decline in modularity at times corresponding to the emergence or transformation of the area.
In superstring theory, the decline aligns with this documented transition and provides an external reference point for the framework.
A further analysis of higher-order network science shows that this decline reflects a cross-disciplinary integration.
In graph representation learning, whose intellectual base expanded rapidly, the decline is gradual, and the framework detects it as a sequence of large effect sizes across successive time windows; in 2022 and beyond, modularity rises again, which we interpret as a re-differentiation of the intellectual base following the integration of the emergence period.
These results suggest that a decline in the modularity of a co-citation network can serve as a structural signature that retrospectively characterizes the emergence and transformation of research areas accompanied by the integration of the intellectual base.

\section{Methods}

\begin{figure*}[t]
\centering
\includegraphics[width=1.0\textwidth]{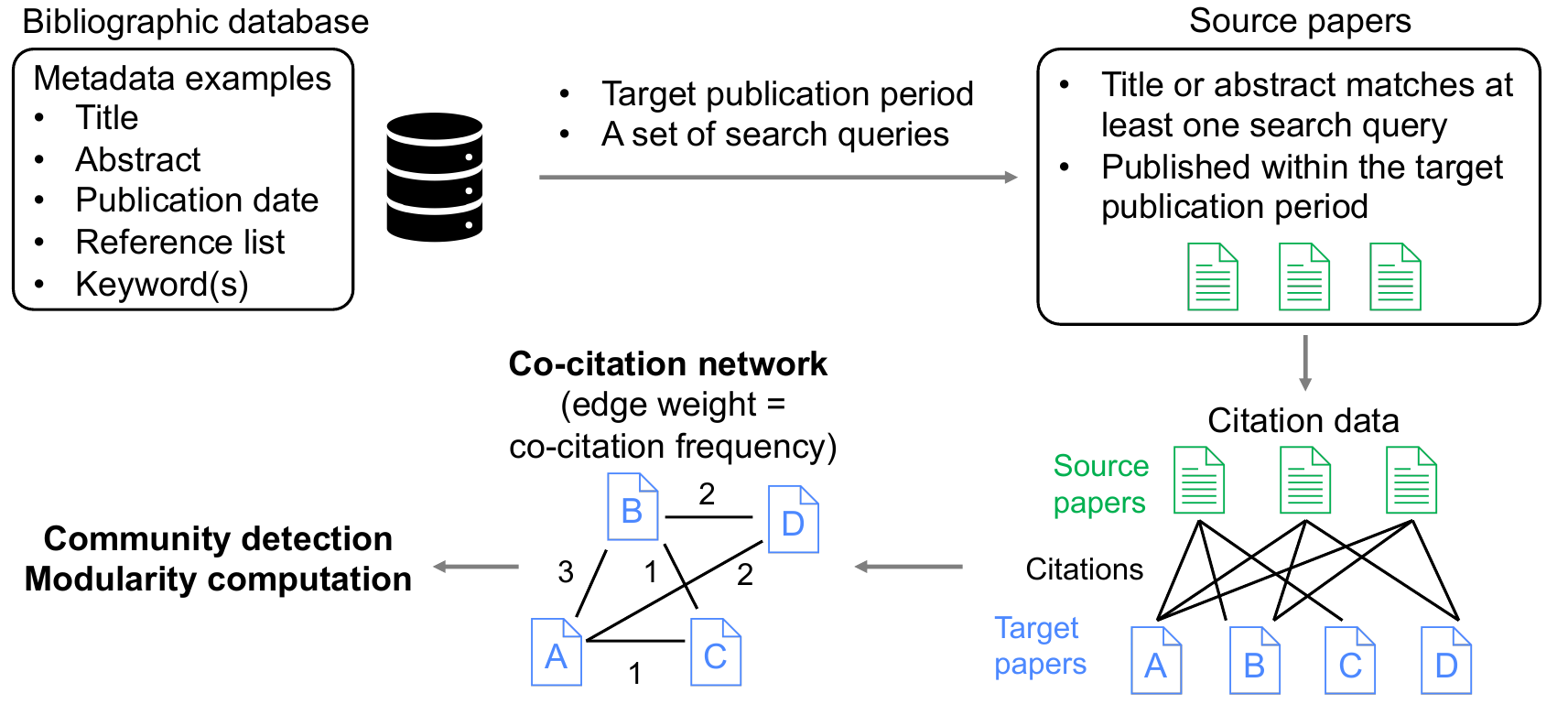}
\caption{
Construction of a weighted co-citation network and computation of modularity from a single set of source papers.
}
\label{fig:1}
\end{figure*}

\begin{figure*}[t]
\centering
\includegraphics[width=1.0\textwidth]{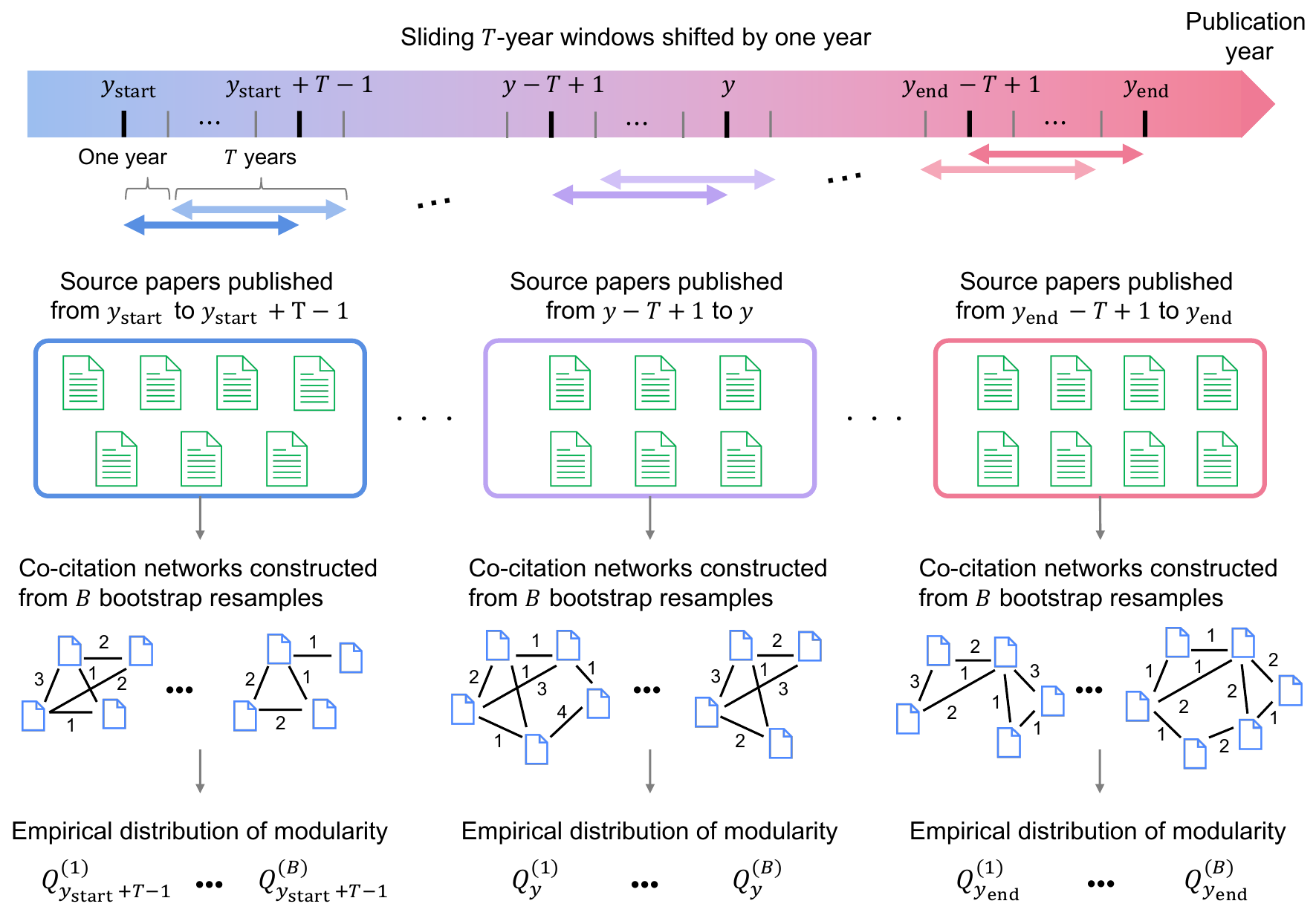}
\caption{
Sliding-window, bootstrap-based time-series analysis of modularity.
The pipeline in Fig.~\ref{fig:1} is applied to overlapping windows of width $T$ years, with a one-year increment.
Within each window, $B$ bootstrap resamples of the source papers yield an empirical distribution of modularity $\{Q_y^{(b)}\}_{b=1}^B$ for $y = y_{\text{start}}, \ldots, y_{\text{end}}$.
}
\label{fig:2}
\end{figure*}

The goal of this study is to quantitatively capture the dynamic structural changes in the intellectual base that occur as a new research area emerges.
We focus on community structure, the modular organization of the intellectual base, and present a framework for analyzing the temporal evolution of modularity, a measure of the strength of that structure.
Figures~\ref{fig:1} and \ref{fig:2} together provide an overview of the framework: Figure~\ref{fig:1} shows how a weighted co-citation network and its modularity are obtained from a single source-paper set, and Figure~\ref{fig:2} shows the sliding-window, bootstrap-based analysis of how this modularity evolves over time.

\subsection{Data}

We use two bibliographic databases depending on the research area under analysis: the cross-disciplinary OpenAlex database~\citep{priem2022} (public snapshot of September 30, 2025) and the INSPIRE HEP database~\citep{inspire-hep, inspire-hep-snapshot}, which specializes in high-energy physics (snapshot of January 8, 2021). 
From each database, we use the title, abstract, publication date, reference list, and document type of each paper.
We also use the editor-assigned INSPIRE keywords for the INSPIRE HEP database.

For each research area, we collect the papers whose titles or abstracts (or INSPIRE keywords) match a set of area-specific search queries within a specified publication-year range, restricted to standard document types; we refer to these papers as ``source papers'' and denote their set by $\mathcal{X}$. 
Among the papers cited by papers in $\mathcal{X}$, we refer to those that do not themselves belong to $\mathcal{X}$ as ``target papers'', and we denote their set by $\mathcal{Y}$. 
We exclude papers in $\mathcal{X}$ from the nodes of the co-citation network because this study focuses on the structure of the external intellectual base on which the field draws.
Because the same paper is often recorded under different IDs, for example, as preprint and journal versions, we deduplicate the papers in $\mathcal{X}$ and those in $\mathcal{Y}$, separately, based on title similarity and first-author matching. 
We describe the data sources, the collection procedure, and the deduplication procedure in detail in Supplementary Section S1.
\begin{table*}[t]
\centering
\caption{The three research areas analyzed in this study. For each area, the table lists the data source, the analysis period, and the numbers of source papers and target papers.}
\label{table:1}
\begin{tabular}{llccc}
\toprule
Research area & Data source & Period & Source papers & Target papers \\
\midrule
Higher-order network science  & OpenAlex    & 2000--2024 & 16,530 & 136,355 \\
Superstring theory            & INSPIRE HEP & 1975--2005 & 24,275 &  51,000 \\
Graph representation learning & OpenAlex    & 2005--2024 & 45,433 & 363,288 \\
\bottomrule
\end{tabular}
\end{table*}

Table~\ref{table:1} summarizes the data source, analysis period, and numbers of source and target papers for each research area.
Supplementary Fig.~S1 shows the time evolution of the number of source papers in each area.

\subsection{Time-series analysis of co-citation networks}

We first describe the co-citation network analysis for a single set of source papers $\mathcal{U} \subseteq \mathcal{X}$ published within a specified period. We then conduct the time-series analysis by applying this procedure across a series of time windows.

\subsubsection{Co-citation network}
We define the co-citation frequency $w_{ij}$ of a paper pair $(v_i, v_j) \in \mathcal{Y} \times \mathcal{Y}$ as the number of source papers in $\mathcal{U}$ that include both $v_i \in \mathcal{Y}$ and $v_j \in \mathcal{Y}$ in the same reference list.
For $i = j$, we set $w_{ii} = 0$.
We construct a co-citation network in which every paper pair satisfying a co-citation frequency $w_{ij} \geq 1$ forms an edge, with the weight of edge $(v_i, v_j)$ given by $w_{ij}$.
While a co-citation network can be disconnected, we take its largest connected component as the object of analysis, which reflects the principal intellectual base~\citep{shibata2008, shibata2009}.
We denote the numbers of nodes and edges of the largest connected component by $N$ and $M$, respectively, and its set of nodes by $\mathcal{V} = \{v_1, \ldots, v_N\} \subseteq \mathcal{Y}$.
We define the adjacency matrix $\mathbf{A} = (A_{ij})_{1 \leq i, j \leq N}$ as the $N \times N$ symmetric matrix
\begin{align}
A_{ij} =
\begin{cases}
1 & (w_{ij} \geq 1) \\
0 & (\text{otherwise})
\end{cases}
\label{eq:1}
\end{align}
We represent the weighted co-citation network $\mathcal{G}$ as the pair of the adjacency matrix $\mathbf{A}$ and the edge weights $\mathbf{w} = (w_{ij})_{1 \leq i, j \leq N}$.

\subsubsection{Modularity}
We use modularity~\citep{newman2004,newman2006} as a measure of the strength of the community structure of a network.
Modularity quantifies the extent to which the edges of a network are concentrated within communities, relative to a null model of a random network that preserves the degree of each node.
A value of $Q = 0$ indicates an edge distribution equivalent to that of the null model.
A higher modularity means that nodes within the same community are more densely connected, whereas nodes in different communities are more sparsely connected.

Consider partitioning the node set $\mathcal{V}$ of the network into $K$ mutually disjoint subsets (communities).
Letting node $v_i \in \mathcal{V}$ belong to community $c_i \in \{1, \ldots, K\}$, the community partition of the entire network is defined as the assignment of community labels to all nodes, $\bm{c} = \{c_1, c_2, \ldots, c_N\}$.
We define the modularity of a weighted network as follows~\citep{newman2004}:
\begin{align}
Q = \frac{1}{2W} \sum_{i=1}^N \sum_{j=1}^N \left( w_{ij} - \frac{k_i^{\text{w}} k_j^{\text{w}}}{2W} \right) \delta(c_i, c_j)
\label{eq:2}
\end{align}
where $W = \sum_{i=1}^N \sum_{j=1}^N w_{ij} / 2$ is the total edge weight, $k_i^{\text{w}} = \sum_{j=1}^N w_{ij}$ is the weighted degree of node $v_i$, and $\delta$ is the Kronecker delta.

\subsubsection{Community detection}

The stochastic block model~\citep{holland1983, fienberg1985, anderson1992, faust1992} is a generative model of random graphs in which a set of $N$ nodes is partitioned into $K$ mutually disjoint communities.
We perform community detection on the weighted co-citation network $\mathcal{G} = (\mathbf{A}, \mathbf{w})$ using a weighted extension~\citep{peixoto2018} of the degree-corrected stochastic block model (DCSBM)~\citep{karrer2011}, which accounts for the heterogeneity of node degrees within communities.
The edge weights are modeled by a discrete Poisson distribution in its microcanonical formulation~\citep{peixoto2014, peixoto2018}.
We estimate the community partition $\bm{c}$ and the number of communities $K$ under the minimum description length (MDL) criterion~\citep{peixoto2013}, which determines $K$ automatically while avoiding overfitting.
The inference proceeds in two stages~\citep{peixoto2014}: a greedy agglomerative procedure yields an initial partition, and an MCMC method that stochastically reassigns nodes to blocks refines it.
We used the \texttt{minimize\_blockmodel\_dl} function of the \texttt{graph-tool} library~\citep{graph-tool} with \texttt{deg\_corr=True}, \texttt{recs=[weight]}, and \texttt{rec\_types=["discrete-poisson"]}.

\subsubsection{Empirical distribution of modularity via bootstrapping}

The modularity of a co-citation network constructed from a single set of source papers $\mathcal{U}$ is a point estimate and may be sensitive to the particular composition of $\mathcal{U}$.
To assess this uncertainty, we introduce a bootstrap procedure that resamples the source papers \citep{rosvall2010, masuda2016}.
From the original set of source papers $\mathcal{U}$, we draw $|\mathcal{U}|$ source papers with replacement to generate a bootstrap sample as a multiset.
We then construct a co-citation network in which the co-citation frequency $w'_{ij}$ is the number of source papers in the bootstrap sample (counted with multiplicity) whose reference lists contain both $v_i \in \mathcal{V}$ and $v_j \in \mathcal{V}$ with $i \neq j$.
We repeat this sampling procedure $B = 10^3$ times to obtain $B$ bootstrap samples $\mathcal{U}^{(1)}, \ldots, \mathcal{U}^{(B)}$.
From these, we obtain the corresponding set of weighted co-citation networks $\{\mathcal{G}'_1, \ldots, \mathcal{G}'_B\}$, where $\mathcal{G}'_b$ is constructed from the $b$-th bootstrap sample $\mathcal{U}^{(b)}$.
For each bootstrap co-citation network $\mathcal{G}'_b$, we again extracted its largest connected component before community detection and the modularity computation.
By computing the modularity $Q^{(b)}$ of each network $\mathcal{G}'_b$, we obtain the empirical distribution of modularity $\{Q^{(1)}, \ldots, Q^{(B)}\}$.

\subsubsection{Contribution of source papers to modularity}

Based on the $B$ bootstrap samples, we quantify the extent to which each source paper is associated with a decline in the modularity of the co-citation network $\mathcal{G}$.
Let $m_u^{(b)}$ denote the multiplicity (i.e., the number of times drawn) of source paper $u$ in the bootstrap sample $\mathcal{U}^{(b)}$.
We partition the sample indices $\{1, \ldots, B\}$ into two sets: $\mathcal{B}^{\mathrm{in}}_u = \{b \mid m_u^{(b)} \geq 1\}$, the indices of the bootstrap samples in which $u$ is drawn at least once, and $\mathcal{B}^{\mathrm{out}}_u = \{b \mid m_u^{(b)} = 0\}$, those in which $u$ is never drawn.
The contribution score is based on the presence or absence of $u$, not on its multiplicity.
We define the contribution score of source paper $u$ as Cliff's $\delta$~\citep{cliff1993, cliff2014} between the modularity distribution $\{Q^{(b)}\}_{b \in \mathcal{B}^{\mathrm{out}}_u}$ over the samples not containing $u$ and the modularity distribution $\{Q^{(b')}\}_{b' \in \mathcal{B}^{\mathrm{in}}_u}$ over the samples containing $u$:
\begin{align}
\delta_u = \frac{
 \left|\{(b,b') \in \mathcal{B}^{\mathrm{out}}_u \times \mathcal{B}^{\mathrm{in}}_u : Q^{(b)} > Q^{(b')}\}\right|
 - \left|\{(b,b') \in \mathcal{B}^{\mathrm{out}}_u \times \mathcal{B}^{\mathrm{in}}_u : Q^{(b)} < Q^{(b')}\}\right|
}{|\mathcal{B}^{\mathrm{out}}_u|\,|\mathcal{B}^{\mathrm{in}}_u|}
\label{eq:9}
\end{align}
The score $\delta_u$ is a nonparametric effect size that takes values in $[-1, 1]$, which is consistent with the fact that $Q$ is bounded in $[0, 1]$ and cannot be assumed to be normally distributed.
A value of $\delta_u > 0$ means that the bootstrap samples containing source paper $u$ tend to have lower modularity than those not containing it, and $\delta_u = 1$ means that every sample containing $u$ has lower modularity than every sample not containing it.
We adopt $\delta_u \geq 0.47$ as the criterion for a large effect size~\citep{romano2006, meissel2025}.
Under this criterion, a source paper is regarded as strongly associated with the modularity decline only when the bootstrap samples containing it have lower modularity than those not containing it in roughly three-quarters or more of all pairwise comparisons (in the absence of ties).

\subsubsection{Sliding-window time-series analysis}

To capture the dynamic changes in the modularity of co-citation networks, we conduct a time-series analysis using a sliding-window approach.
Specifically, we set time windows of width $T$ years with a one-year increment and construct, for each window, the set of source papers published between year $y-T+1$ and year $y$.
For every time window over the entire analysis period from $y_{\text{start}}$ to $y_{\text{end}}$, we perform the community detection, modularity computation, and bootstrap-based statistical evaluation described in the preceding sections.

Unless we state otherwise, we set the window width to $T = 3$ years in the analyses of all research areas.
We confirmed that the main modularity declines do not depend on this choice by comparing the results for $T = 2, 3, 4$ years (Supplementary Fig.~S2).
We assign each source paper to time windows according to its publication date as recorded in the database used for the analysis; for some papers in the INSPIRE HEP database in particular, this date may reflect the preprint release year rather than the publication year of the later journal version.
All publication years and citation counts reported in this study are those recorded in the database snapshots described above.

The DCSBM inference on weighted networks requires constructing a co-citation network and performing the MCMC optimization for each bootstrap sample, and the computational cost of a single MCMC run is $O(M)$~\citep{peixoto2014}.
We run an independent DCSBM inference for each of the $B = 10^3$ bootstrap samples.
In our analysis, we parallelized the computation to reduce the running time.

The representative value of modularity for each time window and its 95\% bootstrap interval are obtained as the mean and the 2.5th and 97.5th percentiles of the empirical distribution, respectively.
This interval represents the sensitivity of modularity with respect to the resampling of the source-paper set.
Note that this variability stems primarily from the resampling of the source papers rather than from the stochasticity of the DCSBM inference (see Supplementary Section S5).

We quantify the change in the empirical distribution of modularity between consecutive time windows.
As with $\delta_u$, we use the nonparametric effect size Cliff's $\delta$~\citep{cliff1993, cliff2014}.
For the empirical distribution $\{Q_y^{(i)}\}_{i=1}^{B}$ of the time window $[y-T+1, y]$ and the empirical distribution $\{Q_{y+1}^{(j)}\}_{j=1}^{B}$ of the time window $[y-T+2, y+1]$, both of width $T$ years, Cliff's $\delta$ is defined as
\begin{align}
  \delta_y = \frac{\left|\left\{(i,j): Q_y^{(i)} > Q_{y+1}^{(j)}\right\}\right| - \left|\left\{(i,j): Q_y^{(i)} < Q_{y+1}^{(j)}\right\}\right|}{B^2}
\label{eq:10}
\end{align}
and takes values in $[-1, 1]$.
A value of $\delta_y > 0$ indicates that the modularity of the co-citation network in the time window $[y-T+1, y]$ tends to be higher than that in the time window $[y-T+2, y+1]$; conversely, $\delta_y < 0$ indicates that it tends to be lower.
As with $\delta_u$, we adopt $\delta_y \geq 0.47$ as the criterion for a large effect size~\citep{romano2006, meissel2025}.
Note that adjacent time windows of width $T = 3$ years share two years of source papers; we therefore use $\delta_y$ as an effect size that quantifies the magnitude of the change between the two distributions, not as a formal hypothesis test, which would presuppose their independence.

\section{Results}

\subsection{Analyzed research areas}

To test whether a decline in the modularity of co-citation networks can serve as an indicator of the emergence and transformation of research areas, we selected three research areas that differ in era, discipline, and growth pattern (Supplementary Fig.~S1).
First, we examine higher-order network science as an emerging area that has been taking shape rapidly in recent years.
Second, we examine superstring theory as a mature area with known historical turning points.
Third, we examine graph representation learning as an area in machine learning that grew rapidly in the late 2010s.

Higher-order network science is a research area of network science that aims to accurately describe the structure and dynamics of complex systems by representing systems involving interactions among three or more units as hypergraphs or simplicial complexes~\citep{battiston2020}.
Such higher-order dependencies beyond pairwise interactions arise in many real-world complex systems, including conversations in social groups~\citep{stehle2011}, collaborations in research~\citep{patania2017}, protein interactions~\citep{wong2008}, and functional brain activity~\citep{giusti2016}.
Since around 2020, this area has attracted rapidly growing attention on both theoretical and applied fronts~\citep{battiston2021, boccaletti2023}.
It therefore provides a suitable case for examining how the integration of an intellectual base unfolds in an emerging area that is still taking shape.

Superstring theory is a research area of theoretical physics that, by modeling elementary particles as one-dimensional strings, seeks to reconcile general relativity with quantum mechanics and to arrive at a Theory of Everything~\citep{green20121, green20122}.
It developed from the 1970s to the early 1980s against the background of related work on string theory, supergravity, and field theory.
The discovery of anomaly cancellation by Green and Schwarz in 1984~\citep{green1984anomaly} then triggered the first superstring revolution, and superstring theory rapidly attracted attention as a candidate theory that treats particle physics and quantum gravity in a unified manner.
Superstring theory is thus a mature area with a well-documented historical turning point~\citep{rickles2014}, providing a suitable case for testing whether the measures used in this study can detect this known transition.

Graph representation learning is a research area that applies machine-learning methods to graph-structured data to learn low-dimensional vector representations of nodes, edges, or entire graphs~\citep{hamilton2020}.
It developed on methodological foundations that include spectral graph theory~\citep{vonluxburg2007}, graph kernels~\citep{vishwanathan2010}, semi-supervised learning~\citep{zhu2003}, and network embedding~\citep{perozzi2014}.
In particular, after 2017, the advent of graph neural networks, exemplified by graph convolutional networks~\citep{kipf2017}, rapidly advanced the fusion of deep learning and graph data analysis.
As a result, graph representation learning has grown rapidly in machine learning and has been deployed in diverse application domains.
Graph representation learning thus offers an appropriate case for analyzing structural changes in an intellectual base, as an area that developed through the integration of multiple methodological lineages.

\subsection{Declining modularity of the intellectual bases}

\begin{figure*}[p]
\centering
\includegraphics[width=1.0\textwidth]{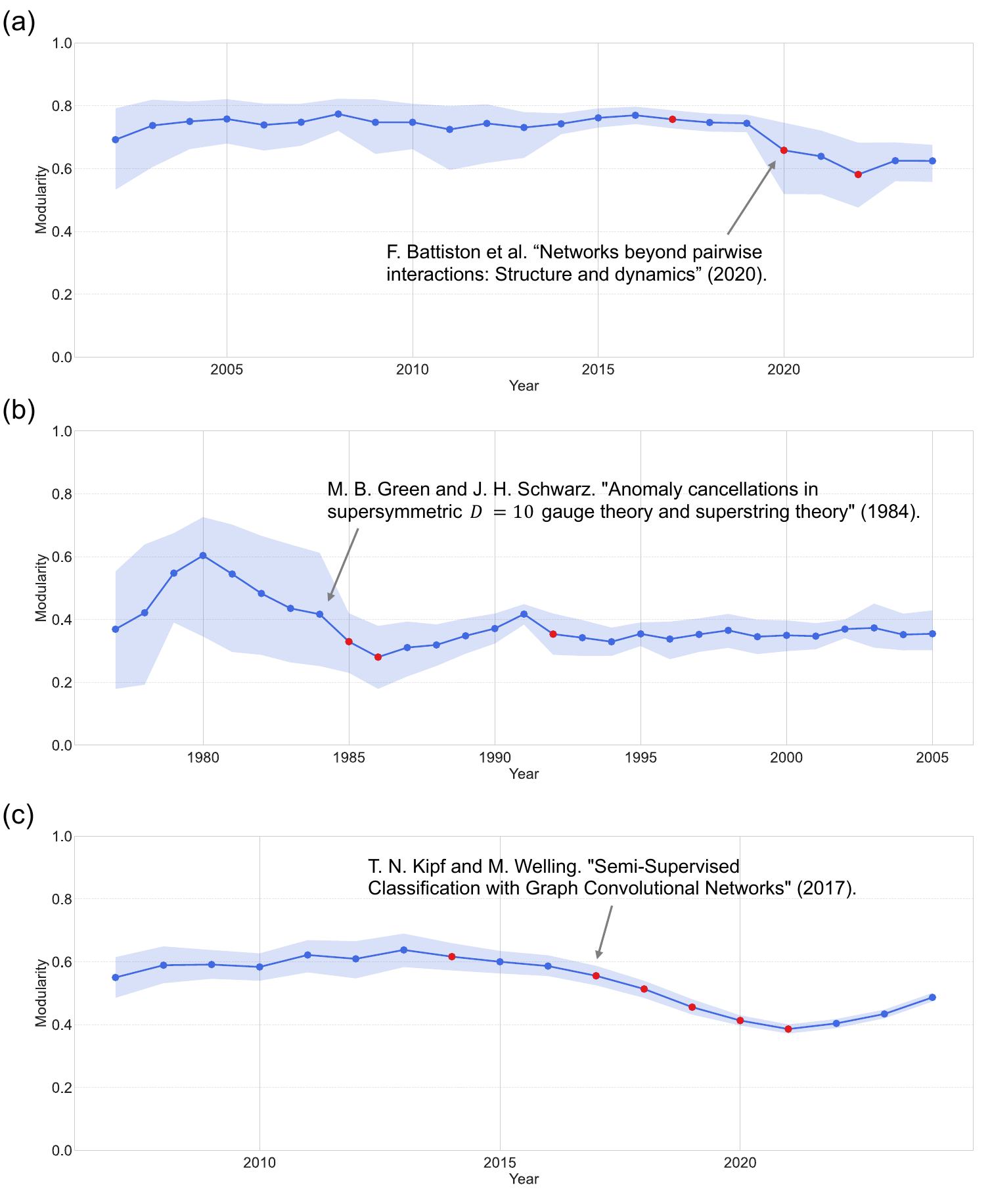}
\caption{Time evolution of the modularity of the intellectual bases in the three research areas. (a) Higher-order network science. (b) Superstring theory. (c) Graph representation learning. Solid lines show the means over 1,000 bootstrap samples, and shaded areas show the 95\% bootstrap intervals. Each three-year time window is plotted at its final year; for example, the value at year 2020 in (a) corresponds to the time window 2018--2020. Red points mark the time windows in which modularity declined from the preceding time window with a large effect size ($\delta \geq 0.47$; see Methods). Annotations indicate representative papers associated with the structural changes in each area.}
\label{fig:3}
\end{figure*}

\begin{figure*}[p]
\centering
\includegraphics[width=1.0\textwidth]{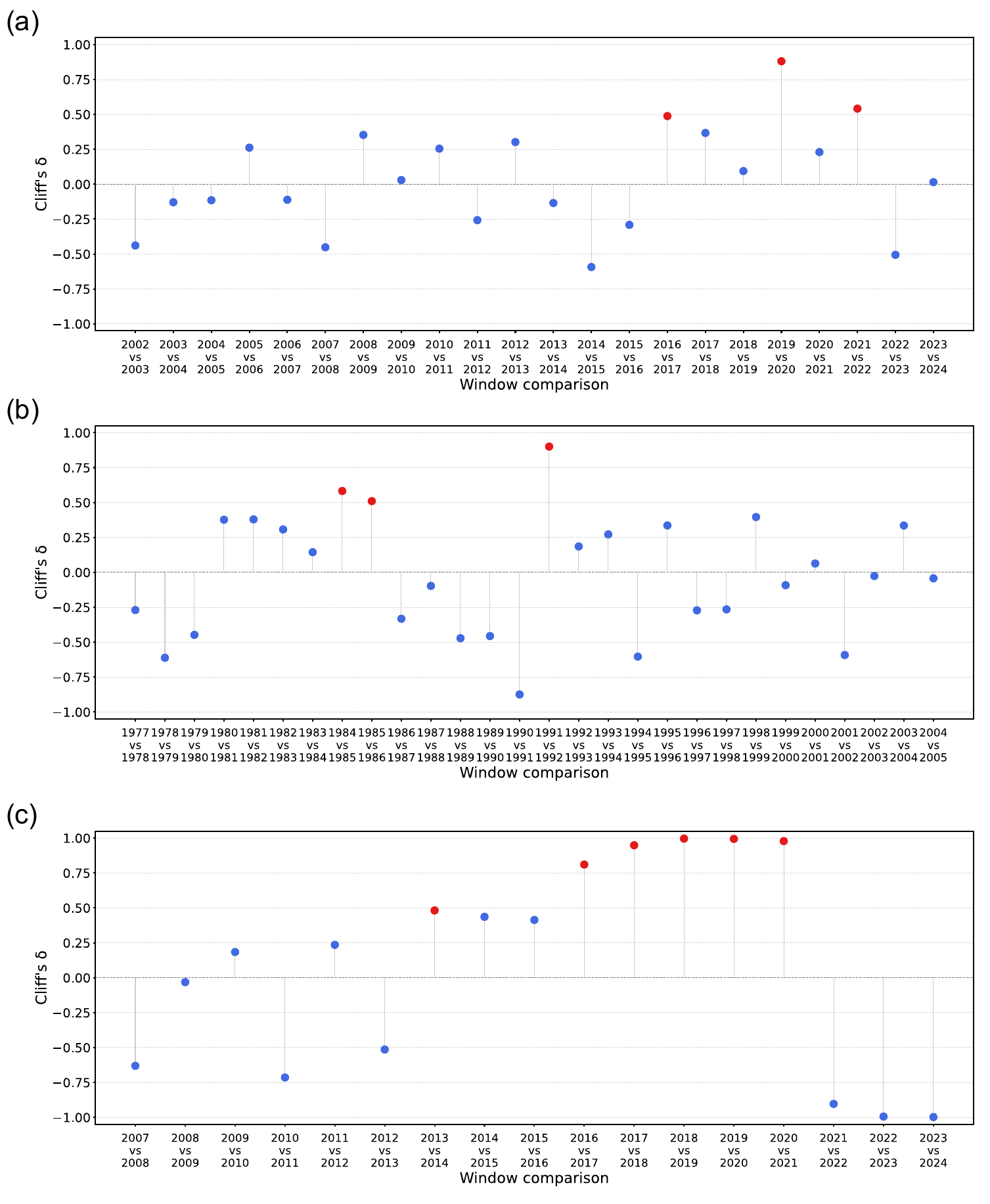}
\caption{Changes in the modularity distribution between adjacent time windows. (a) Higher-order network science. (b) Superstring theory. (c) Graph representation learning. Each pair of adjacent time windows is labeled by their final years; for example, ``2019 vs 2020'' compares the time windows 2017--2019 and 2018--2020. Cliff's $\delta$ quantifies the change in the modularity distribution from one time window to the next, with positive values indicating a decline in modularity. Red points indicate the time windows satisfying $\delta \geq 0.47$, the criterion for a large effect size; these are the same time windows marked red in Fig.~\ref{fig:3}.}
\label{fig:4}
\end{figure*}

\begin{figure*}[t]
\centering
\includegraphics[width=1.0\textwidth]{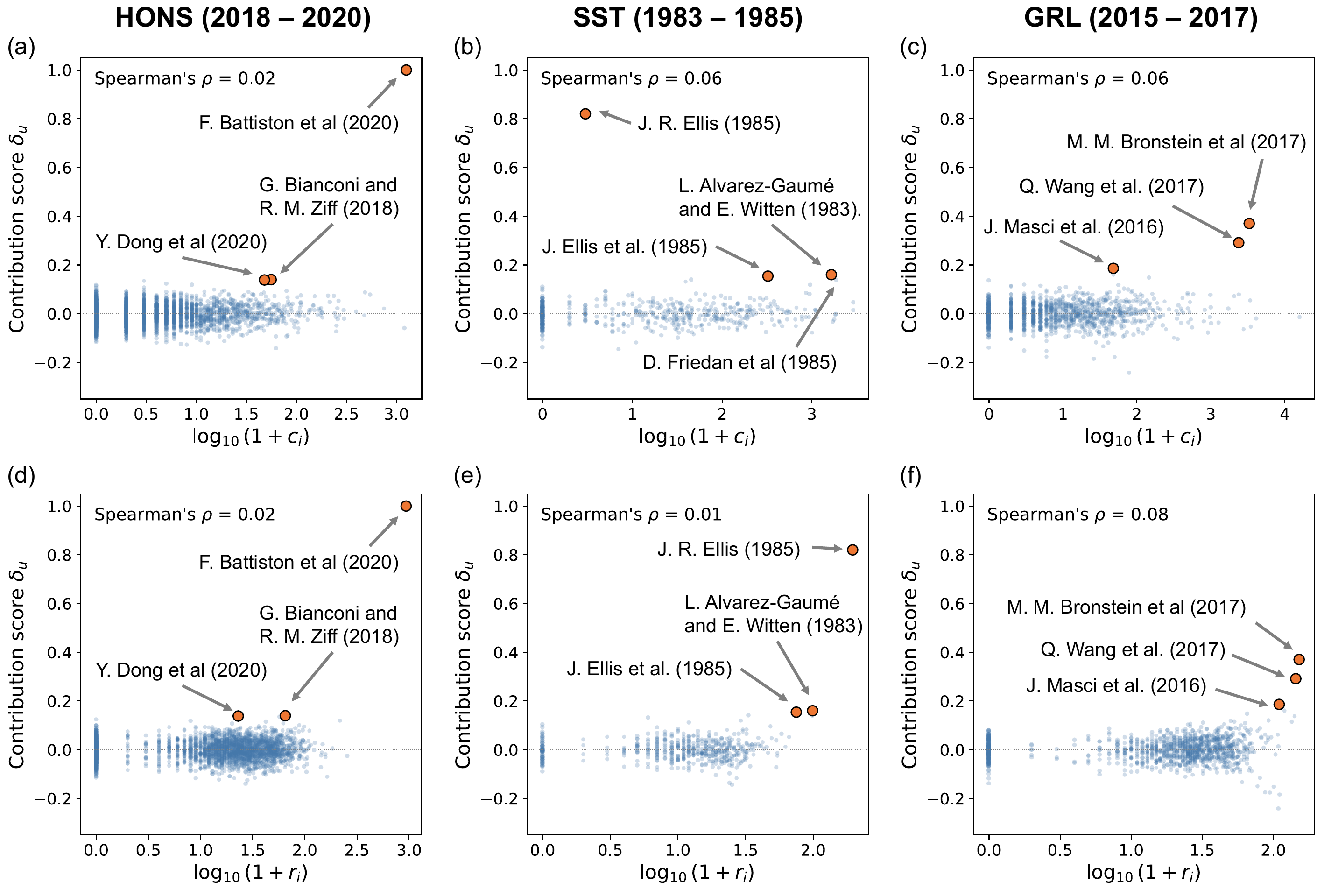}
\caption{Contribution scores $\delta_u$ of the source papers in the analyzed time window of each research area. The top row (a)--(c) plots $\delta_u$ against the number of incoming citations, and the bottom row (d)--(f) against the number of outgoing references. (a) and (d): Higher-order network science (2018--2020). (b) and (e): Superstring theory (1983--1985). (c) and (f): Graph representation learning (2015--2017). In each panel, the three source papers with the highest $\delta_u$ are annotated, and Spearman's rank correlation coefficient $\rho$ is shown.}
\label{fig:5}
\end{figure*}

To quantify structural changes in the intellectual base of each research area, we constructed a co-citation network for each three-year sliding window and analyzed the time evolution of its modularity.
In these networks, nodes represent target papers cited by the source papers, and edges represent pairs of target papers co-cited by the same source paper.
High modularity indicates that co-citations are concentrated within largely separate groups of target papers; we interpret this structure as an intellectual base divided into distinct research communities.
A decline in modularity, in turn, indicates that co-citations increasingly connect target papers across previously separate groups; we interpret this change as an integration of the intellectual base.
We therefore hypothesize that the modularity of a co-citation network declines during the formation of a new research frontier, as papers that bridge existing knowledge domains appear.

Figure~\ref{fig:3} shows the time evolution of modularity in the three research areas.
We found that in all three areas the modularity decline aligns with a period in which existing research communities became connected in a new research context, whereas the temporal pattern of the decline differs from area to area.

In higher-order network science, modularity remained high from the 2000s through the late 2010s and then declined markedly in the time window ending in 2020 (2018--2020).
The decline coincides with the publication of the review article by Battiston et al.~\citey{battiston2020}; we interpret this as indicating that theoretical and applied research on higher-order interactions began to be connected to a common intellectual base~\citep{battiston2021, majhi2022, boccaletti2023}.

In superstring theory, modularity rose in the late 1970s and then declined rapidly from the early to the mid-1980s.
This suggests that multiple research communities that had been forming in the field's early years were integrated within a short period.
The decline corresponds to the first superstring revolution, triggered by the discovery of anomaly cancellation by Green and Schwarz~\citey{green1984anomaly}; we interpret this as reflecting the process by which the previously separate intellectual bases of particle physics, gravity, and field theory became integrated around string theory~\citep{green20121, green20122}.

In graph representation learning, modularity was relatively high from the late 2000s to the early 2010s and then declined continuously from 2016 through 2021.
This suggests that the integration of different research topics in the intellectual base proceeded over multiple years.
We interpret the decline as reflecting the process by which, after the graph convolutional network paper by Kipf and Welling~\citey{kipf2017}, the intellectual bases of spectral graph theory, semi-supervised learning, and neural networks developed into research on graph neural networks~\citep{hamilton2020}.
Modularity reached its minimum in the 2019--2021 window and then rose monotonically through the 2022--2024 window.
A rise in modularity indicates that co-citations increasingly concentrate within separate groups of target papers; we interpret this as a re-differentiation of the intellectual base following the integration of the emergence period.

Thus, modularity declines capture not a single mode of research-area formation but several distinct ones: integration led by a review article (higher-order network science), integration driven by a theoretical revolution (superstring theory), and the convergence of methodological foundations (graph representation learning).

The time evolution of modularity shown in Fig.~\ref{fig:3} is the bootstrap mean in each time window, and it is difficult to judge from point estimates alone whether a change reflects a genuine shift of the modularity distribution or sampling variation.
In particular, when a modularity decline proceeds gradually over multiple years, the change between adjacent time windows appears small.
We therefore compare the bootstrap empirical distributions of modularity in adjacent time windows and assess how consistently the distribution as a whole has shifted.
For this comparison, we use the nonparametric effect size Cliff's $\delta$~\citep{cliff1993, cliff2014} (see Methods).

Figure~\ref{fig:4} shows the changes in the modularity distribution between adjacent time windows in terms of Cliff's $\delta$; under our definition, positive values indicate a decline in modularity (see Methods).
Adopting $\delta \geq 0.47$ as the criterion for a large effect size~\citep{romano2006, meissel2025}, we find that large effect sizes appear in each area in the periods expected as its transition periods.
Specifically, marked modularity declines were detected in 2020 in higher-order network science, in 1985 (and 1986) in superstring theory, and from 2017 through 2021 in graph representation learning.
These timings coincide with well-documented developments in each area: the publication of the review article in higher-order network science, the first superstring revolution in superstring theory, and the rise of graph neural networks in graph representation learning.
In graph representation learning in particular, although the point estimates in Fig.~\ref{fig:3} begin to drift downward around 2014, the decline with large effect sizes is sustained from 2017 through 2021, indicating that a gradual yet robust integration proceeded over multiple years.
In 2022 and beyond, large effect sizes appear in the opposite direction ($\delta = -0.90$, $-0.99$, and $-1.00$ for the three adjacent-window comparisons after the 2019--2021 window; Fig.~\ref{fig:4}(c)), indicating that the rise in modularity is as consistent across bootstrap samples as the preceding decline.

In each area, declines with large effect sizes were also observed in time windows relatively distant from these main transition periods.
Local declines appeared in 2017 in higher-order network science and in 2014 in graph representation learning; these can be interpreted as early structural changes that precede the main integrations.
In superstring theory, a large effect size was also observed in 1992; this may reflect the reorganization of the intellectual base around conformal field theory and two-dimensional quantum gravity in the aftermath of the first superstring revolution~\citep{difrancesco1995}.
Similarly, in higher-order network science, a large effect size was observed in 2022, following the main decline in 2020, suggesting that the integration continued for some years beyond the initial transition.
These results indicate that the proposed measures may capture not only the main transition of an area but also multiple structural changes, including precursory changes and subsequent reorganizations.
Taken together, they also confirm that the modularity declines observed in Fig.~\ref{fig:3} do not rest on point estimates alone but represent consistent shifts of the entire bootstrap distributions.

We next examine to what extent individual source papers are associated with these declines.
For each area, we computed the contribution score $\delta_u$ of each source paper in the time window in which the marked decline coincided with the documented development (higher-order network science: 2018--2020; superstring theory: 1983--1985; graph representation learning: 2015--2017).
The score $\delta_u$ is larger when the bootstrap samples containing a given source paper tend to have lower modularity; it thus quantifies the extent to which the paper bridges different knowledge communities and thereby corresponds to the modularity decline (see Methods).

Figure~\ref{fig:5} plots $\delta_u$ against two bibliometric quantities of each source paper, namely the number of incoming citations and the number of outgoing references, and annotates the three source papers with the highest $\delta_u$ in each panel.
The two quantities correspond to two alternative explanations of a high score: if $\delta_u$ merely reflected the visibility of a paper, it would correlate with the number of incoming citations, and if it merely reflected the length of a reference list, which mechanically produces many co-citation pairs, it would correlate with the number of outgoing references.
In all areas, Spearman's rank correlation coefficient was below 0.1 for both quantities.
$\delta_u$ is therefore not explained by such quantitative prominence; instead, it reflects the degree to which a paper's references bridge multiple knowledge communities.

In higher-order network science, the review article by Battiston et al.~\citey{battiston2020} showed an exceptionally high $\delta_u$ in the 2018--2020 time window ($\delta_u > 0.99$), far exceeding all other papers.
This suggests that the review, which organized theories, models, data analysis, and applications of higher-order interactions across previously separate lines of research, functioned as a single bridge connecting their intellectual bases to a common research context.

In superstring theory, papers from the first superstring revolution period showed high $\delta_u$ in the 1983--1985 time window, including those on gravitational anomalies~\citep{alvarez1984gravitational} and on superstrings and conformal invariance~\citep{friedan1986conformal}.
Notably, the highest $\delta_u$ was recorded by the lecture notes of J.~Ellis~\citey{ellis1985supersymmetry}, which broadly surveyed particle phenomenology while receiving only a few citations (Fig.~\ref{fig:5}(b,~e)); this is a clear example of a bridging role that citation counts cannot capture.
This is consistent with a picture in which the intellectual bases of anomalies, supersymmetry, supergravity, and string phenomenology became connected in the same period, in the wake of the discovery of anomaly cancellation by Green and Schwarz~\citey{green1984anomaly}.

In graph representation learning, multiple papers showed high $\delta_u$ in the 2015--2017 time window, including those on geometric deep learning~\citep{bronstein2017geometric} and a survey of knowledge graph embedding~\citep{wang2017knowledge}.
This suggests that the intellectual base of graph representation learning formed not around a single representative paper but through the convergence of multiple methodological streams spanning spectral methods, representation learning, knowledge graphs, and deep learning.

This contrast can also be confirmed at the level of the modularity decline itself.
For each of the time windows analyzed above, we split the bootstrap samples into those containing the source paper with the highest $\delta_u$ and those not containing it, and compared each subset with the preceding time window.
The fraction of the modularity decline that remains after excluding that paper was smallest in higher-order network science (roughly 20\%) and larger in superstring theory and graph representation learning (roughly 50\% and 80\%, respectively; see Supplementary Section S3).
That is, the decline in higher-order network science depends strongly on the single review~\citep{battiston2020}, whereas the declines in superstring theory and graph representation learning largely persist even without the most strongly associated paper, indicating a more collective character.

These analyses revealed a different mode of integration in each area.
In higher-order network science, a single review served as the dominant bridge, whereas in superstring theory and graph representation learning, we observed a collective integration in which multiple papers played comparable bridging roles.

\subsection{Cross-disciplinary integration of the intellectual base in higher-order network science}

\begin{figure*}[p]
\centering
\includegraphics[width=0.9\textwidth]{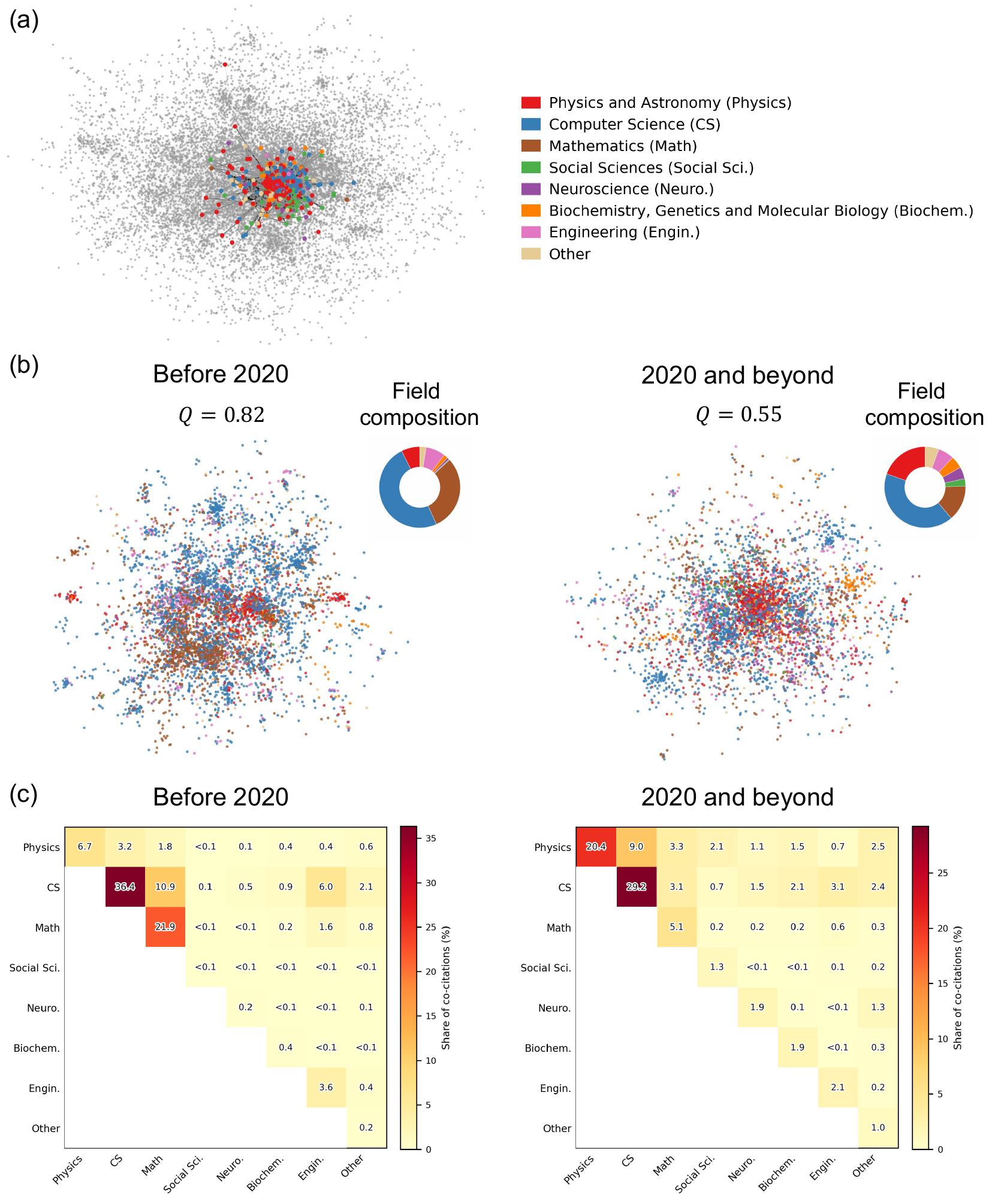}
\caption{Cross-disciplinary integration of the intellectual base in higher-order network science around 2020.
(a) Co-citation pattern generated by Battiston et al.~\citey{battiston2020}. Colored nodes represent the target papers included in the reference list of this single source paper, edges represent co-citations between them, and gray nodes represent the other target papers in the co-citation network. The background co-citation network is built from all source papers over the full analysis period (2000--2024), using the same construction as in panel~(b).
(b) Comparison of the co-citation networks before 2020 and in 2020 and beyond. Each network shows the largest connected component consisting of edges with a co-citation frequency of at least three. Nodes represent target papers, and the donut charts show their field composition.
(c) Shares of co-citations within and between fields, before 2020 and in 2020 and beyond.
In panels (a) and (b), node colors indicate fields in the OpenAlex classification.
Each target paper is assigned to the field of its OpenAlex primary topic; papers whose field is outside the seven displayed fields, and papers without a primary topic, are grouped as ``Other.''}
\label{fig:6}
\end{figure*}

\begin{table*}[p]
\caption{Representative frequently co-cited paper pairs in higher-order network science.}
\label{table:3}
\begin{center}
\begin{tabular}{cX}
\toprule
\multicolumn{2}{l}{\textbf{Before 2020}} \\
\midrule
Times co-cited & Co-cited paper pair \\
\midrule
\multirow{2}{*}{75} & \labelitemi\ L.~Qi. ``Eigenvalues of a real supersymmetric tensor'' (2005). \\
 & \labelitemi\ L.-H.~Lim. ``Singular values and eigenvalues of tensors: A variational approach'' (2005). \\ \midrule
\multirow{2}{*}{70} & \labelitemi\ B.~W.~Kernighan and S.~Lin. ``An efficient heuristic procedure for partitioning graphs'' (1970). \\
 & \labelitemi\ C.~M.~Fiduccia and R.~M.~Mattheyses. ``A linear-time heuristic for improving network partitions'' (1988). \\ \midrule
\multirow{2}{*}{49} & \labelitemi\ K.-C.~Chang, K.~Pearson, and T.~Zhang. ``Perron-frobenius theorem for nonnegative tensors'' (2008). \\
 & \labelitemi\ Q.~Yang and Y.~Yang. ``Further results for perron-frobenius theorem for nonnegative tensors II'' (2011). \\ \midrule
\multirow{2}{*}{44} & \labelitemi\ H.~Edelsbrunner, D.~Letscher, and A.~Zomorodian. ``Topological persistence and simplification'' (2002). \\
 & \labelitemi\ A.~Zomorodian and G.~Carlsson. ``Computing persistent homology'' (2004). \\ \midrule
\multirow{2}{*}{35} & \labelitemi\ D.~J.~Watts and S.~H.~Strogatz. ``Collective dynamics of 'small-world' networks'' (1998). \\
 & \labelitemi\ A.-L.~Barab{\'a}si and R.~Albert. ``Emergence of scaling in random networks'' (1999). \\
\midrule
\multicolumn{2}{l}{\textbf{2020 and beyond}} \\
\midrule
Times co-cited & Co-cited paper pair \\
\midrule
\multirow{2}{*}{100} & \labelitemi\ T.~N.~Kipf and M~Welling. ``Semi-supervised classification with graph convolutional networks'' (2017). \\
 & \labelitemi\ P.~Veli{\v{c}}kovi{\'c} et al. ``Graph attention networks'' (2018). \\ \midrule
\multirow{2}{*}{57} & \labelitemi\ D.~J.~Watts and S.~H.~Strogatz. ``Collective dynamics of 'small-world' networks'' (1998). \\
 & \labelitemi\ A.-L.~Barab{\'a}si and R.~Albert. ``Emergence of scaling in random networks'' (1999). \\ \midrule
\multirow{2}{*}{51} & \labelitemi\ L.~Qi. ``Eigenvalues of a real supersymmetric tensor'' (2005). \\
 & \labelitemi\ L.-H.~Lim. ``Singular values and eigenvalues of tensors: A variational approach'' (2005). \\ \midrule
\multirow{2}{*}{41} & \labelitemi\ J.~Stehl{\'e}. ``High-resolution measurements of face-to-face contact patterns in a primary school'' (2011). \\
 & \labelitemi\ R.~Mastrandrea et al. ``Contact patterns in a high school: a comparison between data collected using wearable sensors, contact diaries and friendship surveys'' (2015). \\ \midrule
\multirow{2}{*}{41} & \labelitemi\ G.~Petri et al. ``Homological scaffolds of brain functional networks'' (2014). \\
 & \labelitemi\ F.~Battiston et al. ``The physics of higher-order interactions in complex systems'' (2021). \\
\bottomrule
\end{tabular}
\end{center}
\end{table*}

Focusing on higher-order network science, we now descriptively characterize how the modularity decline observed around 2020 manifested in the intellectual base.
We set the boundary at 2020 because the sliding-window analysis showed the largest Cliff's $\delta$ in the 2018--2020 window and the review by Battiston et al.~\citey{battiston2020}, the source paper most strongly associated with that decline, was published in 2020.

Figure~\ref{fig:6}(a) shows the co-citation pattern generated by the review paper~\citep{battiston2020}.
As a review spanning theories, models, data analysis, and applications of higher-order interactions, this paper cites papers from multiple fields, including physics, mathematics, computer science, and neuroscience, within a single reference list.
This suggests that the review by Battiston et al. (2020) was not merely a highly cited paper but functioned as a bridging paper that connected research belonging to previously separate intellectual bases to the common framework of higher-order network science.

To examine how this bridging corresponds to structural changes in the intellectual base as a whole, we split the set of source papers into those published before 2020 and those published in 2020 and beyond, and constructed a static co-citation network for each period (Fig.~\ref{fig:6}(b)).
Here, we visualize the largest connected component consisting of co-citations supported by at least three source papers.
The network before 2020 consists of 5,474 target papers and 76 communities, with a modularity of $Q = 0.820$.
In contrast, the network in 2020 and beyond consists of 4,469 target papers and 67 communities, and its modularity dropped to $Q = 0.549$.
These results confirm that the modularity decline observed in the sliding-window analysis is reproduced in a static before-and-after comparison, and that the intellectual base of the area shifted from a state divided into multiple communities to a more integrated one in 2020 and beyond.
The field composition of the target papers, shown by the donut charts in Fig.~\ref{fig:6}(b), also changed substantially.
Before 2020, target papers in computer science and mathematics accounted for the majority of the network.
In 2020 and beyond, the share of mathematics shrank considerably while that of physics rose markedly, and applied fields such as neuroscience, biochemistry, and the social sciences appeared, diversifying the field composition.

To further describe how this integration was accompanied by changes with respect to the field classification, we divided the co-citations into those within a single field and those between different fields.
Before 2020, co-citations between different fields accounted for 30.9\% of the total; in 2020 and beyond, this share increased to 37.3\%.
At the same time, assortative mixing by field~\citep{newman2002_2} decreased from 0.533 to 0.493.
Assortative mixing quantifies the tendency for connections to form between elements that share the same attribute; here, it takes higher values when target papers in the same OpenAlex field are more likely to be co-cited.
These results are consistent with the interpretation that the modularity decline in 2020 and beyond was not merely a decrease in the number of communities but was accompanied by more frequent co-citation of target papers from different academic fields by the same source papers.

Figure~\ref{fig:6}(c) shows which within-field and between-field co-citation shares changed across the 2020 boundary.
Before 2020, co-citations within computer science, within mathematics, and between computer science and mathematics accounted for large shares, consistent with an intellectual base rooted in mathematical and computer-science foundations such as tensor analysis, hypergraphs, combinatorial optimization, and graph algorithms.
Around 2020, this composition changed markedly.
The shares of co-citations within computer science and within mathematics decreased, reducing the relative weight of the connections among the mathematical and computer-science topics that had been central to the intellectual base.
Instead, the share of co-citations within physics increased, and the shares between physics and other fields, including computer science, the social sciences, neuroscience, and biochemistry, rose across the board; as a result, multiple applied fields became connected through physics.
This change suggests that the intellectual base of higher-order network science was reorganized from a composition centered on mathematics and computer science to one that centers on physics and incorporates machine learning, social systems, neuroscience, and biological applications.

This field-pair-level change is also visible in the content of the specific paper pairs that are frequently co-cited (Table~\ref{table:3}).
Before 2020, the top co-cited pairs included classic papers on tensor eigenvalue theory, graph partitioning algorithms, the Perron--Frobenius theory of nonnegative tensors, persistent homology, and complex networks.
This is consistent with an intellectual base organized around the theoretical foundations of mathematics, computer science, and complex networks.
In 2020 and beyond, the top co-cited pairs came to include papers on graph neural networks, empirical studies of face-to-face contact patterns, topological analyses of functional brain networks, and the physics of higher-order interactions.
At the same time, the pair of complex-network papers by Watts and Strogatz and by Barab{\'a}si and Albert, and the pair of tensor-eigenvalue papers by Qi and by Lim, remained among the top pairs in 2020 and beyond.
The changes in 2020 and beyond can therefore be interpreted as a process in which machine learning, social systems, neuroscience, and the physics of higher-order interactions became connected on top of the existing theoretical core.

Taken together, these results indicate that the modularity decline in higher-order network science around 2020 can be interpreted not as a mere homogenization of the intellectual base but as a cross-disciplinary integration that accompanied the bridging citation pattern centered on the review by Battiston et al. (2020).
That is, as knowledge from physics, neuroscience, the social sciences, and biological applications became connected to the existing mathematical and computer-science foundations, the body of research on higher-order interactions came to be visible as a single new research area.

\section{Discussion}

In this study, we focused on the temporal evolution of the modularity of co-citation networks and proposed a framework for quantifying the structural changes in the intellectual base that accompany the emergence of research areas.
Through the application to three research areas, higher-order network science, superstring theory, and graph representation learning, our results suggested that the emergence of a new research area is observed as a decline in the modularity of the co-citation network, which we interpret as an integration of knowledge communities that had previously been largely separate.
The comparison of the three research areas showed that the modularity decline takes a different form in each area: review-led integration in higher-order network science, integration through a theoretical revolution in superstring theory, and the convergence of methodological streams in graph representation learning.

We extend a line of bibliometric research that, through co-citation analysis, addresses which research communities or research fronts exist and how their composition changes over time~\citep{small1973, boyack2010, shibata2008, shibata2009, trujillo2018, chen2006, rosvall2010}.
We focus on how the strength of the community structure of the intellectual base changes as a research area emerges.
As a closely related prior work, Shwed and Bearman~\citey{shwed2010} tracked the modularity of citation networks over time and interpreted its decline as the formation of scientific consensus on contested propositions.
While our framework shares with theirs the idea of reading a decline in modularity as an epistemic integration, it differs in three respects: it targets the emergence of research areas rather than the resolution of scientific controversies; it analyzes the co-citation structure of the external intellectual base rather than direct citations within the focal literature; and it evaluates modularity as a bootstrap distribution with effect sizes and paper-level contribution scores rather than as a point estimate.

The observed declines are largely robust to the choice of the community-detection algorithm.
While the main analyses detect communities with the DCSBM, we also repeated the bootstrap analysis with the Leiden algorithm, which maximizes modularity directly~\citep{traag2019} (see Supplementary Section S6 for details).
Under the Leiden algorithm, the declines at the main transitions and the subsequent rise in graph representation learning were reproduced, and the source paper with the largest contribution score was identical in every area, although the sensitivity of the two algorithms differs in data-sparse periods such as the mid-1980s in superstring theory.
This agreement indicates that the detected integration reflects the structure of the co-citation data rather than the choice of the inference method.
It is worth mentioning that the Leiden variant also runs up to two orders of magnitude faster per bootstrap sample in our environment.
For corpora larger than those analyzed in this study, the framework can therefore be applied with the Leiden algorithm and validated with the DCSBM on selected time windows.

Modularity can depend on network size (the number of nodes $N$ and edges $M$).
Community detection by modularity maximization is known to suffer from a resolution limit, whereby communities smaller than a scale determined by the total number of edges cannot be detected~\citep{fortunato2007, fortunato2010}; the DCSBM inference we employed under the minimum description length criterion is not directly subject to this limit~\citep{peixoto2014}.
Nevertheless, in our data, modularity showed weak-to-moderate negative correlations with $N$ and $M$ across time windows.
Indeed, the Spearman rank correlation coefficients were $-0.42$ for $N$ and $-0.43$ for $M$ in higher-order network science, $-0.39$ for $N$ and $-0.43$ for $M$ in superstring theory, and $-0.69$ for both $N$ and $M$ in graph representation learning.
We thus cannot rule out the possibility that network growth driven by the increasing number of source papers partially affects the long-term trend of modularity.
The central claims of this study, however, rest on the distributional changes between adjacent time windows (Cliff's $\delta$) and on the declines in specific transition periods.
Moreover, in the before-and-after comparison of higher-order network science around 2020, modularity dropped from 0.820 to 0.549 even though the network shrank from 5,474 to 4,469 nodes, in the opposite direction to what the network-size correlation would imply.
Similarly, in graph representation learning, modularity rose monotonically after the 2019--2021 window even though the network more than doubled in size, again opposite to the direction implied by the negative size correlation.
We therefore interpret the observed declines as reflecting substantive structural changes in the intellectual base rather than changes in network size.

The structural variation analysis by Chen~\citey{chen2012} quantifies the degree to which a newly published paper, through its reference list, connects previously distinct clusters in an existing co-citation network.
Chen measured this boundary spanning with metrics including the modularity change rate and cluster linkage, and showed that cluster linkage in particular is a better predictor of the paper's future citation counts than commonly studied variables such as the number of cited references.
Our contribution score $\delta_u$ shares the idea of quantifying the association between an individual source paper and the structure of the intellectual base.
The two approaches differ, however, in two respects.
First, whereas structural variation analysis measures structural change as a point estimate on a single network, $\delta_u$ is defined as an effect size (Cliff's $\delta$) between bootstrap empirical distributions of modularity and incorporates sampling uncertainty.
Second, whereas structural variation analysis primarily aims to predict future citation counts, our $\delta_u$ aims to describe the integration process of the intellectual base itself, independently of citation counts (Fig.~\ref{fig:5}).
In this sense, the two approaches are complementary: structural variation analysis asks which papers will attract future attention, whereas $\delta_u$ characterizes how the intellectual base reorganizes as a research area emerges.

The interpretation of the contribution score $\delta_u$ requires caution.
Firstly, a small $\delta_u$ for an individual source paper does not necessarily imply that the paper played no part in the integration: when many source papers redundantly bridge the same knowledge communities, the presence or absence of any single one may hardly change the modularity, and all of their $\delta_u$ values can therefore be small.
In contrast, an exceptionally large $\delta_u$ identifies a paper that acts as a dominant, largely non-redundant bridge.
Secondly, although $\delta_u$ was almost uncorrelated with the number of references in all three areas (Spearman's $\rho < 0.1$; see Fig.~\ref{fig:5}), a review, which usually has a long reference list, can both homogenize the intellectual base and be a substantive integration event in the formation of a research area.
On one hand, McMahan and McFarland~\citey{mcmahan2021} showed, using a large-scale corpus, that the publication of a review reorganizes the attention structure of the literature.
On the other hand, generative artificial intelligence recently makes review-like articles with long reference lists easier to produce~\citep{elazar2026, smyth2026}.
Requiring co-citations to be supported by multiple independent source papers, as in the thresholded construction we used for higher-order network science (Supplementary Section S3), can help separate such artificial homogenization from substantive integration.
Moreover, $\delta_u$ captures a statistical association between the presence of a source paper and modularity, not a causal effect.
A paper with a large $\delta_u$ may have prompted the integration, or may have appeared in response to an integration already under way---a review, for instance, may be written precisely because an area has begun to coalesce.
The phrase ``review-led integration'' should therefore be read as describing a structural role (i.e., the bridging of multiple knowledge communities by a reference list) rather than a causal one.

Several additional limitations could affect the analysis.
First, the set of source papers depends on the search queries, and because the boundaries of a research area are not uniquely defined, a different query design could alter the modularity trajectory.
For superstring theory in particular, we mitigated the limited coverage of older publications by also using the INSPIRE subject classification terms (tag~695).
Second, the document-type classifications and recorded publication dates in the databases are imperfect (e.g., atypical documents may remain after filtering, and preprint dates can shift window assignments by up to a few years).
The aggregate trajectories, based on thousands of source papers and a three-year window, are expected to mitigate the effects of such errors.
Third, the completeness of the recorded reference lists varies over time: in our OpenAlex snapshot, the fraction of source papers with no recorded references is relatively large in recent years due to the indexing lag (e.g., 48.7\% in higher-order network science and 40.8\% in graph representation learning in 2024).
Thus, the most recent results, such as the rise of modularity in graph representation learning in 2022 and beyond, should be carefully interpreted.
Finally, our evidence is limited in scope: we analyzed three areas whose emergence or transformation is known in retrospect, so the case selection is itself a form of confirmation, though within these areas the large effect sizes were concentrated in the expected transition windows rather than spread across all time windows (Fig.~\ref{fig:4}).
Moreover, the emergence of a research area may not involve integrating the intellectual base; P{\"a}{\"a}kk{\"o}nen et al.~\citey{paakkonen2026} reported that the author co-citation network of computational social science grew more fragmented from 2000 to 2020.
Our results should therefore be read as capturing an integrative mode of emergence, and prospective validation and applications to areas that may remain fragmented are left for future work.

Despite these limitations, by focusing on the dynamic decline of the modularity of the intellectual base, we proposed that the emergence of a research area can be captured as an integration of previously largely separate knowledge communities.
Although we constructed the set of source papers for each research area by manually designing search queries, existing subject classifications, such as the topics in OpenAlex~\citep{priem2022}, or corpora anchored to individual review articles may reduce this manual effort and enable systematic comparisons across a large number of research areas.
We expect that applications to a broader range of research areas and prospective validation on emerging areas will advance our understanding of the dynamics of science and technology and will clarify the extent to which research frontiers can be detected prospectively.

\section*{Acknowledgments}
K.N. acknowledges the financial support from JST ACT-X Grant Number JPMJAX24CI.
K.N. and Y.S. acknowledge the financial support from JST ASPIRE Grant Number JPMJAP2328.
K.N. and M.A. acknowledge the financial support from JSPS KAKENHI Grant Number JP25H01122.

\section*{Declaration of Competing Interest}
The authors declare no competing interests.

\section*{Data Availability}
This study uses two openly available bibliographic databases: OpenAlex~\citep{priem2022} (public snapshot of September 30, 2025) and INSPIRE HEP~\citep{inspire-hep, inspire-hep-snapshot} (snapshot of January 8, 2021).
The derived datasets essential for reproducing the results are publicly available on Zenodo~\citep{nakajima2026dataset}.

\section*{Code Availability}
The code essential for reproducing the results is publicly available at \url{https://github.com/kazuibasou/cocitation-modularity}.

%% file: sm.tex
\begin{center}
\vspace*{12pt}
{\Large Supplementary Information for:\\
\vspace{12pt}
Declining Modularity of Intellectual Bases During the Emergence of Research Areas}
\vspace{12pt} \\
\end{center}

\setcounter{figure}{0}
\setcounter{table}{0}
\setcounter{section}{0}
\setcounter{equation}{0}

\renewcommand{\thesection}{S\arabic{section}}
\renewcommand{\thefigure}{S\arabic{figure}}
\renewcommand{\thetable}{S\arabic{table}}
\renewcommand{\theequation}{S\arabic{equation}}

\section{Data collection and preprocessing}

\subsection{Data sources}

We use two bibliographic databases depending on the research area under analysis.

For the analyses of higher-order network science and graph representation learning, we use the OpenAlex database~\citep{priem2022}, which indexes papers with rich bibliographic metadata across disciplines.
From the OpenAlex public snapshot of September 30, 2025, we extracted 220,432,395 papers published between 1950 and 2025.
For each paper, the following bibliographic information is available:
(i) title, (ii) abstract, (iii) publication date, (iv) reference list, and (v) type.
In this study, we analyze papers whose OpenAlex \texttt{type} is one of ``article'', ``preprint'', ``review'', or ``book-chapter''.
Although ``book'' is an important document type, we excluded it because its records frequently had missing or non-substantive abstracts (e.g., preface text) and incomplete reference lists.

For the analysis of superstring theory, we use the INSPIRE HEP database~\citep{inspire-hep}, which indexes papers in particle physics and related fields, specializing in high-energy physics.
Maintained through continuous, internationally coordinated curation by major institutions such as CERN, DESY, and Fermilab, it provides high-quality bibliographic metadata, including reference lists, for high-energy physics papers from the 1970s onward.
From the snapshot of January 8, 2021~\citep{inspire-hep-snapshot}, we extracted 1,381,017 papers published between 1950 and 2020.
For each paper, the following bibliographic information is available:
(i) title, (ii) abstract, (iii) publication date, (iv) reference list, (v) collection type, and (vi) keywords (MARC tag 695; the ``INSPIRE keywords'' field, a controlled vocabulary specific to INSPIRE assigned by editors).
In this study, we analyze papers whose INSPIRE collection type is one of ``HEP'', ``REVIEW'', or ``PROCEEDINGS''.
For the same reason as with OpenAlex, we excluded the ``BOOK'' collection type from the analysis.

\subsection{Collection of source papers}

We constructed the set of source papers for each research area by designing a set of search queries $\mathcal{S}_{\text{query}}$ relevant to the area and collecting the papers that match them.
Each query is matched case-insensitively using regular-expression matching that respects word boundaries.
Each element of $\mathcal{S}_{\text{query}}$ takes one of two forms:
(i) a single-keyword condition (the title or abstract contains the keyword as a whole word); or
(ii) an AND condition (the title or abstract contains every constituent keyword as a whole word).
For each keyword, we included its singular and plural forms where applicable, together with hyphenation and spelling variants.
Abstracts are available for 52.1\% of papers in OpenAlex and for 74.9\% of papers in INSPIRE HEP.
For papers without an abstract, only the title is searched.

When using OpenAlex, we collect source papers by specifying a publication-year range from $y_{\text{start}}$ to $y_{\text{end}}$ together with the query set $\mathcal{S}_{\text{query}}$.
When using INSPIRE HEP, in addition to the publication-year range and the query set, we also search the INSPIRE-specific keyword field (MARC tag 695).
That is, a paper qualifies as a source paper if its title, abstract, or INSPIRE keywords satisfy any of the query conditions.

The complete list of queries and tag~695 keywords for each area is provided in \texttt{search\_keywords.xlsx}, which is available at Ref.~\citep{nakajima2026dataset}; below we describe the design for each area.

\subsubsection{Higher-order network science}

For higher-order network science, we set $y_{\text{start}} = 2000$ and $y_{\text{end}} = 2024$, and designed the queries with reference to the review article by Battiston et al.~\citey{battiston2020}.
We used single-keyword conditions targeting the core representations of the area, namely `hypergraph', `simplicial complex', and `higher-order network', together with AND conditions that pair `higher-order interaction' with `network'.
This gave 16 queries in total, provided in the ``HONS'' sheet of \texttt{search\_keywords.xlsx}.

\subsubsection{Superstring theory}

For superstring theory, we set $y_{\text{start}} = 1975$ and $y_{\text{end}} = 2005$, and designed the queries with reference to the books by Green, Schwarz, and Witten~\citey{green20121, green20122}.
We use INSPIRE HEP rather than OpenAlex for this area because the key papers in superstring theory are concentrated in the period from 1970 to 2000, for which INSPIRE HEP provides continuously curated bibliographic records including reference lists, and because the editor-assigned INSPIRE keywords (MARC tag~695) complement text-based search queries in collecting source papers.
For INSPIRE HEP, we combined a title and abstract text search with a tag~695 keyword search, restricting the collection types to \texttt{HEP}, \texttt{REVIEW}, and \texttt{PROCEEDINGS}.

For the text search, we used AND conditions that pair spelling variants of `supergravity', `supersymmetry', or `conformal field theory' with `string' or `superstring', together with single keywords for named concepts such as `string theory', `heterotic string', `m-theory', `d-brane', `Nambu-Goto', and `Polyakov action'.
This gave 78 queries in total, provided in the ``SST\_text\_search'' sheet of \texttt{search\_keywords.xlsx}.

For the tag~695 search, we performed an exact-match search using 556 INSPIRE subject classification terms.
Table~\ref{table:s_sst_tag695} summarizes the main categories; the complete list is provided in the ``SST\_tag695'' sheet of \texttt{search\_keywords.xlsx}.

\begin{table}[h]
\centering
\small
\caption{Main categories of the tag~695 keywords for superstring theory.}
\label{table:s_sst_tag695}
\begin{tabular}{lrl}
\toprule
Category & Count & Representative examples \\
\midrule
string model & 188 & \texttt{string model: heterotic}, \texttt{string model: d-brane}, \texttt{string model: calabi-yau} \\
membrane model & 109 & \texttt{membrane model}, \texttt{membrane model: action}, \texttt{membrane model: brst} \\
orbifold & 85 & \texttt{orbifold}, \texttt{orbifold: z(2)}, \texttt{orbifold: calabi-yau} \\
d-brane & 19 & \texttt{d-brane}, \texttt{d-brane: boundary condition}, \texttt{d-brane: charge} \\
m-theory & 15 & \texttt{m-theory}, \texttt{m-theory: compactification}, \texttt{m-theory: matrix model} \\
superstring & 4 & \texttt{superstring}, \texttt{superstring: heterotic} \\
Other & 136 & \texttt{ads/cft correspondence}, \texttt{field theory: string}, \texttt{dimension: 10} \\
\midrule
Total & 556 & \\
\bottomrule
\end{tabular}
\end{table}

\subsubsection{Graph representation learning}

For graph representation learning, we set $y_{\text{start}} = 2005$ and $y_{\text{end}} = 2024$, and designed the queries with reference to the book by Hamilton~\citey{hamilton2020}.
We used three groups of conditions: (i) three-keyword AND conditions combining `spectral', `graph', and one of `neural network', `embedding', or `clustering'; (ii) single keywords for representation- and embedding-based methods, such as `graph kernel', `laplacian eigenmap', `graph embedding', and `node embedding'; and (iii) single keywords for graph neural network methods and their named variants, such as `graph convolutional network', `graph attention network', `graph autoencoder', and `geometric deep learning'.
This gave 34 queries in total, provided in the ``GRL'' sheet of \texttt{search\_keywords.xlsx}.

\subsection{Deduplication of source and target papers}

We denote by $\mathcal{X}$ the set of all source papers over the entire period.
Among the papers cited by papers in $\mathcal{X}$, those that do not themselves belong to $\mathcal{X}$ are called ``target papers'', and we denote the set of all target papers over the entire period by $\mathcal{Y}$.

In bibliographic databases, the same paper is often recorded under different IDs because of title variations, the coexistence of preprint and published versions, and similar factors.
We therefore deduplicate the papers in $\mathcal{X}$ and in $\mathcal{Y}$ separately.
We denote the set of papers to be deduplicated by $\mathcal{Z}$ (that is, $\mathcal{Z} = \mathcal{X}$ or $\mathcal{Z} = \mathcal{Y}$).
First, for the title string of each paper $z \in \mathcal{Z}$, we pad each word with surrounding whitespace to make word boundaries explicit and decompose it into character $n$-grams of length 2 to 4.
We represent each paper as a vector whose components are the TF-IDF (term frequency--inverse document frequency)~\citep{salton1973} values of these character $n$-grams.
Next, we normalize each vector by its $L^2$ norm and compute cosine similarities. 
To avoid the all-pairs computation over all papers, we use sparse matrix operations to efficiently identify only the paper pairs $(z, z')$ with similarity at least 0.75 as duplicate candidates.
For each candidate pair, we compute the normalized Levenshtein distance $d_{\text{norm}}(z, z') = d_{\text{lev}}(s_z, s_{z'}) / \max(|s_z|, |s_{z'}|)$, where $d_{\text{lev}}(s_z, s_{z'})$ is the Levenshtein distance between the title strings $s_z$ and $s_{z'}$, and $|s_z|$ is the length of the string $s_z$.
If $d_{\text{norm}}(z, z') \leq 0.05$ and the first author's ``first-name initial plus surname'' matches, we regard $z$ and $z'$ as the same paper.
Pairs judged to be the same paper are merged into a single paper group.
The reference list of each paper group is taken as the union of the reference lists of all papers in the group.
To avoid double counting, the citation count of each paper group is taken as the maximum citation count among the papers in the group.
The publication year of each paper group is taken as the publication year of the paper with the highest citation count.
This is based on the assumption that the most-cited version is likely to correspond to the formally published version.

\subsection{Time evolution of the number of papers in each case}

\begin{figure}[p]
\centering
\includegraphics[width=\linewidth]{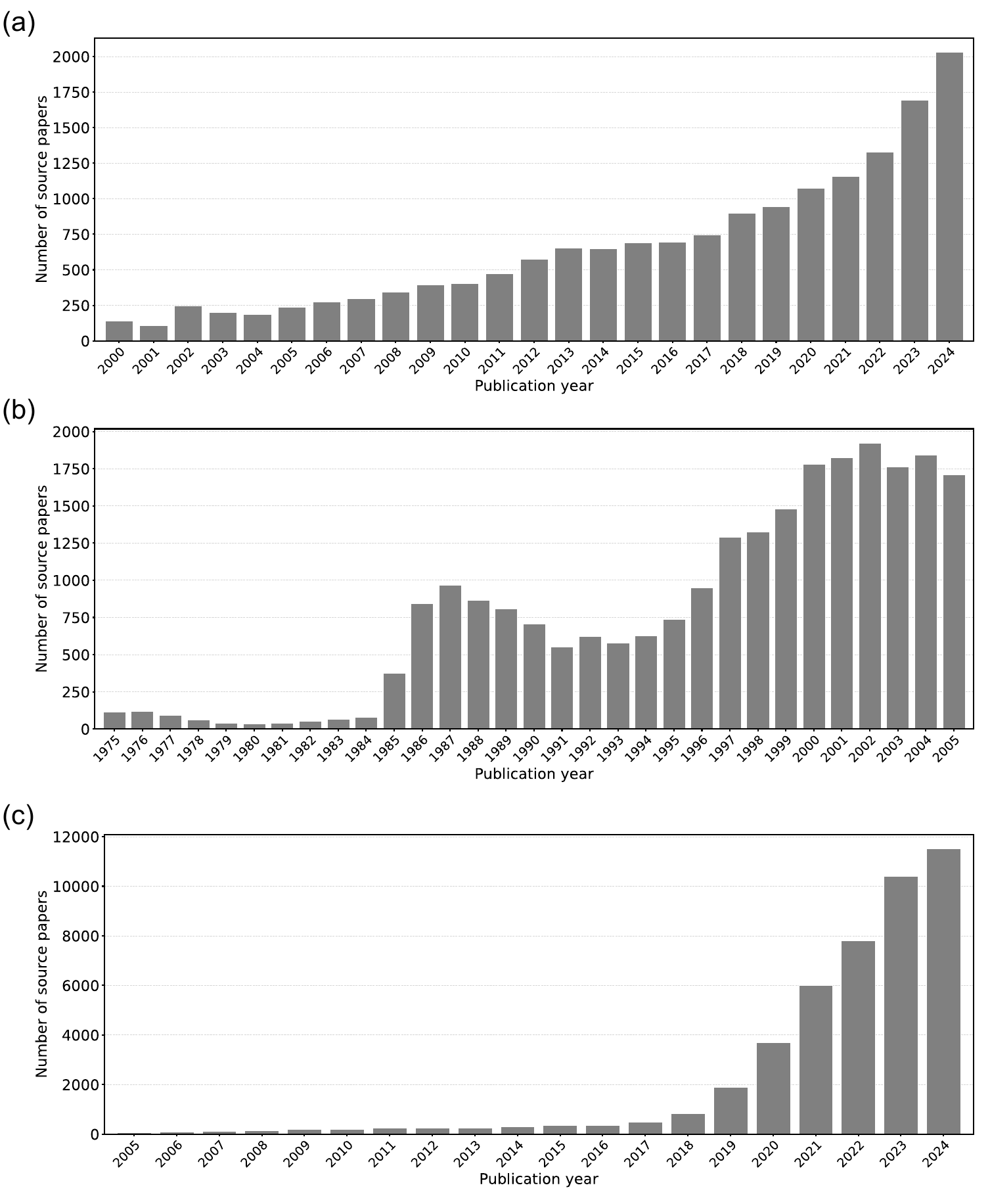}
\caption{Number of source papers per year in each research area. (a) Higher-order network science. (b) Superstring theory. (c) Graph representation learning.}
\label{fig:s1}
\end{figure}

For each research area, Fig.~\ref{fig:s1} shows the number of source papers published in each year.

\section{Unweighted Co-citation Network Analysis}
\label{sec:unweighted}

In this section, we describe the unweighted co-citation network analysis used in the in-depth analysis of higher-order network science in the main text. 
This analysis broadly follows the method of identifying the knowledge structure of a field through co-citation clustering~\citep{trujillo2018}, and it is independent of the sliding-window time-series analysis in the main text, which uses weighted co-citation networks. 
Below, we describe the analysis for a given set of source papers $\mathcal{U} \subseteq \mathcal{X}$ published within a specified period.

\subsection{Co-citation network}

We define the co-citation frequency $w_{ij}$ of a paper pair $(i, j) \in \mathcal{Y} \times \mathcal{Y}$ as the number of source papers in $\mathcal{U}$ whose reference lists contain both papers $i$ and $j$. 
In an unweighted co-citation network, for a given threshold $c$, only the paper pairs with a co-citation frequency $w_{ij} \geq c$ form edges, and no weights are assigned to the edges. 
While a co-citation network is generally disconnected, we take its largest connected component, which reflects the principal intellectual base, as the object of analysis~\citep{shibata2008, shibata2009}. 
We denote the node set of the largest connected component by $\mathcal{V} \subseteq \mathcal{Y}$ and its edge set by $\mathcal{E}$, and we write the numbers of nodes and edges as $N = |\mathcal{V}|$ and $M = |\mathcal{E}|$, respectively.

\subsection{Modularity}

We partition the node set $\mathcal{V}$ of the network into $K$ mutually disjoint communities, with node $i \in \mathcal{V}$ belonging to community $c_i$. 
We define the modularity of an unweighted network as follows~\citep{newman2006}:
\begin{align}
Q = \frac{1}{2M} \sum_{i=1}^N \sum_{j=1}^N \left( A_{ij} - \frac{k_i k_j}{2M} \right) \delta(c_i, c_j)
\end{align}
Here $A_{ij}$ is 1 if there is an edge between nodes $i$ and $j$ and 0 otherwise, $k_i = \sum_{j} A_{ij}$ is the degree of node $i$, and $\delta$ is the Kronecker delta.

\subsection{Selection of the co-citation frequency threshold}

In an unweighted network, the threshold $c$ affects the size and granularity of the network.
To choose an appropriate threshold, we compared $c \in \{1, 3, 5, 7, 9\}$ using three validation criteria adapted from Trujillo et al.~\citey{trujillo2018}.
(i) \textbf{Internal consistency}: the Spearman rank correlation between the degree of each node and the number of source papers in the period that cite the corresponding paper.
Papers cited by more source papers are expected to be more central in the co-citation network.
(ii) \textbf{Community validity}: a $\chi^2$ test of independence between the communities detected by the DCSBM and the OpenAlex fields of the papers, together with Cram{\'e}r's $V$ as an effect size.
An association between the two indicates that the communities correspond to established academic fields.
(iii) \textbf{Stability}: the Spearman rank correlation between the degrees of the nodes shared by the network at threshold $c$ and the network at threshold $1$.
It indicates how well the structure of the original network is preserved as the threshold increases.

We note that our implementation differs from that of Trujillo et al.~\citey{trujillo2018} in two respects.
First, whereas they assessed stability against a second network built from a more comprehensive set of search queries \citep{trujillo2018}, we assess it against the $c=1$ network built from the same queries; because the thresholded networks are nested, every node at threshold $c$ also appears in the $c=1$ network, and we therefore report the rank correlation  rather than the number of matching documents.
Second, Cram{\'e}r's $V$ is not part of the original criteria; we add it because, at our network sizes, the $\chi^2$ test rejects independence trivially.

Table~\ref{table:s_threshold} shows the results.
All three criteria support the validity of the networks across thresholds: the internal-consistency correlation is positive and significant at every threshold, the $\chi^2$ test indicates a clear association between communities and fields, and the stability correlation is moderately positive throughout.
The criteria alone, however, do not single out one threshold.
At threshold $c = 1$, the network is too large (before 2020, $N \approx 6.6 \times 10^4$ with about 650 communities) and too fine-grained to interpret.
At thresholds $c \geq 5$, the network shrinks sharply (before 2020, from 5{,}474 nodes at $c = 3$ to 1{,}490 at $c = 5$ and 83 at $c = 9$) and many co-citations are lost.
Taking these into account, we adopted $c = 3$ as the smallest threshold that yields an interpretable network size and number of communities.

\begin{table}[t]
\centering
\small
\caption{Validation of the unweighted co-citation networks at each co-citation frequency threshold $c$, using three validation criteria adapted from Trujillo et al.~\citey{trujillo2018}: internal consistency, the Spearman rank correlation between node degree and the number of source papers citing the paper; community validity, a $\chi^2$ test of independence between DCSBM communities and OpenAlex fields, with Cram{\'e}r's $V$ as an effect size; and stability, the Spearman rank correlation between the degrees of the nodes shared with the $c=1$ network. All $P$ values are below $0.001$.}
\label{table:s_threshold}
\begin{tabular}{lrrrrr}
\toprule
 & $c=1$ & $c=3$ & $c=5$ & $c=7$ & $c=9$ \\
\midrule
\multicolumn{6}{l}{\textbf{Before 2020}} \\
Number of nodes & 65{,}615 & 5{,}474 & 1{,}490 & 188 & 83 \\
Number of edges & 1{,}736{,}427 & 27{,}822 & 4{,}605 & 469 & 149 \\
Number of communities & 652 & 76 & 28 & 8 & 3 \\
Modularity & 0.64 & 0.82 & 0.90 & 0.71 & 0.54 \\
Internal consistency & 0.418 & 0.551 & 0.618 & 0.606 & 0.776 \\
Community validity $\chi^2$ & 243{,}822 & 9{,}649 & 2{,}367 & 132 & 51 \\
\quad Degrees of freedom & 16{,}275 & 1{,}350 & 378 & 35 & 6 \\
\quad Cram{\'e}r's $V$ & 0.386 & 0.313 & 0.337 & 0.375 & 0.553 \\
Stability & --- & 0.382 & 0.336 & 0.352 & 0.559 \\
\midrule
\multicolumn{6}{l}{\textbf{2020 and beyond}} \\
Number of nodes & 77{,}721 & 4{,}469 & 1{,}173 & 506 & 287 \\
Number of edges & 2{,}695{,}042 & 27{,}481 & 4{,}744 & 1{,}696 & 858 \\
Number of communities & 882 & 67 & 25 & 13 & 8 \\
Modularity & 0.56 & 0.55 & 0.53 & 0.57 & 0.57 \\
Internal consistency & 0.445 & 0.579 & 0.715 & 0.718 & 0.747 \\
Community validity $\chi^2$ & 356{,}221 & 14{,}782 & 2{,}640 & 841 & 333 \\
\quad Degrees of freedom & 22{,}025 & 1{,}254 & 408 & 168 & 84 \\
\quad Cram{\'e}r's $V$ & 0.428 & 0.417 & 0.364 & 0.372 & 0.407 \\
Stability & --- & 0.408 & 0.454 & 0.475 & 0.481 \\
\bottomrule
\end{tabular}
\end{table}

\subsection{Community detection}

The community detection procedure is the same as in the main text, except that the network is unweighted. 
We infer the community partition with the microcanonical DCSBM~\citep{karrer2011, peixoto2014} by minimizing the description length $\Sigma = S_{\bm{c}} + \mathcal{L}_{\bm{c}}$~\citep{peixoto2013}, where $S_{\bm{c}}$ is the microcanonical entropy and $\mathcal{L}_{\bm{c}}$ is the encoding cost of the community partition, both given in the main text. 
Because the network has no edge weights, the description length omits the edge-weight likelihood term of the weighted formulation~\citep{peixoto2018}. 
The two-stage inference (greedy agglomerative initialization followed by MCMC refinement) is also identical to the main text: we used the \texttt{minimize\_blockmodel\_dl} function of the \texttt{graph-tool} library~\citep{graph-tool} with \texttt{deg\_corr=True} and without an edge-weight covariate.

\section{Robustness of the Modularity Decline to the Exclusion of a Single Source Paper}
\label{sec:loo}

In this section, we assess the extent to which the modularity decline observed in the main text depends on the single source paper most strongly associated with it, and whether it reflects integration that extends beyond that paper's own reference list.
As in the time-series analysis in the main text, we use the weighted co-citation network that retains all co-citations (co-citation frequency $\geq 1$).
For each research area, we consider the time window analyzed for the contribution score $\delta_u$ in the main text (Table~\ref{table:s_loo}).
We partition the $B = 10^3$ bootstrap samples in this window into the set $\mathcal{B}^{\mathrm{in}}$ of samples containing the source paper with the largest $\delta_u$ and the set $\mathcal{B}^{\mathrm{out}}$ of samples not containing it.
We compare the mean modularity over $\mathcal{B}^{\mathrm{out}}$ with the mean modularity in the preceding time window, and define the remaining fraction of the decline as the fraction of the decline from the preceding window to the analyzed window that remains in $\mathcal{B}^{\mathrm{out}}$.
A smaller remaining fraction indicates that the decline depends more strongly on that single paper.
Note that this analysis is not a strict leave-one-out procedure that reconstructs the network; rather, it compares the existing bootstrap samples conditioned on the presence or absence of the paper.

Table~\ref{table:s_loo} shows the results.
In higher-order network science, approximately 20\% of the decline remains after excluding the source paper with the largest $\delta_u$, the review by Battiston et al.~($\delta_u > 0.99$); the decline around 2020 thus depends strongly on this single review.
In contrast, the majority of the decline remains in superstring theory (53\%) and graph representation learning (80\%), which indicates a collective integration that does not strongly depend on any single paper.
Across the three areas, the remaining fraction is ordered inversely to the largest $\delta_u$, and the distinction between single-paper and collective integration read from the $\delta_u$ distributions is thus confirmed at the level of the modularity decline itself.

In the weighted network (co-citation frequency $\geq 1$), the long reference list of the review by Battiston et al.\ forms many co-citation edges (a clique) at once, so this single paper appears as the dominant contributor to the decline; this is consistent with its role as the principal bridge across fields (review-led integration).
To test whether this bridging extends beyond the review's own reference list to the intellectual base as a whole, we use the static network with threshold $\geq 3$ (Section~\ref{sec:unweighted}), which retains only co-citations supported by three or more source papers and thus removes the clique.
There, 85\% of the decline remains even after excluding the same review, indicating that the decline in higher-order network science reflects a substantive, cross-disciplinary integration supported by multiple independent source papers rather than a mechanical artifact of a single reference list.

\begin{table}[t]
\centering
\small
\caption{Single-source exclusion analysis in the time window analyzed for the contribution score $\delta_u$ in each research area, using the weighted co-citation network (co-citation frequency $\geq 1$). The excluded paper is the source paper with the largest $\delta_u$ in the window. The three $Q$ rows show the mean modularity over the bootstrap samples of the preceding window, over all samples of the analyzed window, and over the samples not containing the excluded paper, respectively. The remaining fraction is the ratio of the decline that remains after the exclusion.}
\label{table:s_loo}
\begin{tabular}{lccc}
\toprule
 & \shortstack{Higher-order\\network science} & \shortstack{Superstring\\theory} & \shortstack{Graph representation\\learning} \\
\midrule
Time window & 2018--2020 & 1983--1985 & 2015--2017 \\
Excluded paper & Battiston (2020) & Ellis (1985) & Bronstein (2017) \\
$\delta_u$ of the excluded paper & $> 0.99$ & 0.82 & 0.37 \\
$Q$ for the preceding window & 0.744 & 0.417 & 0.586 \\
$Q$ for the analyzed window & 0.658 & 0.330 & 0.556 \\
$Q$ for the samples without the paper & 0.726 & 0.370 & 0.562 \\
Remaining fraction of the modularity decline & 21\% & 53\% & 80\% \\
\bottomrule
\end{tabular}
\end{table}

\section{Robustness of the Modularity Decline to the Choice of Time-Window Width}
\label{sec:window}

\begin{figure}[p]
\centering
\includegraphics[width=\linewidth]{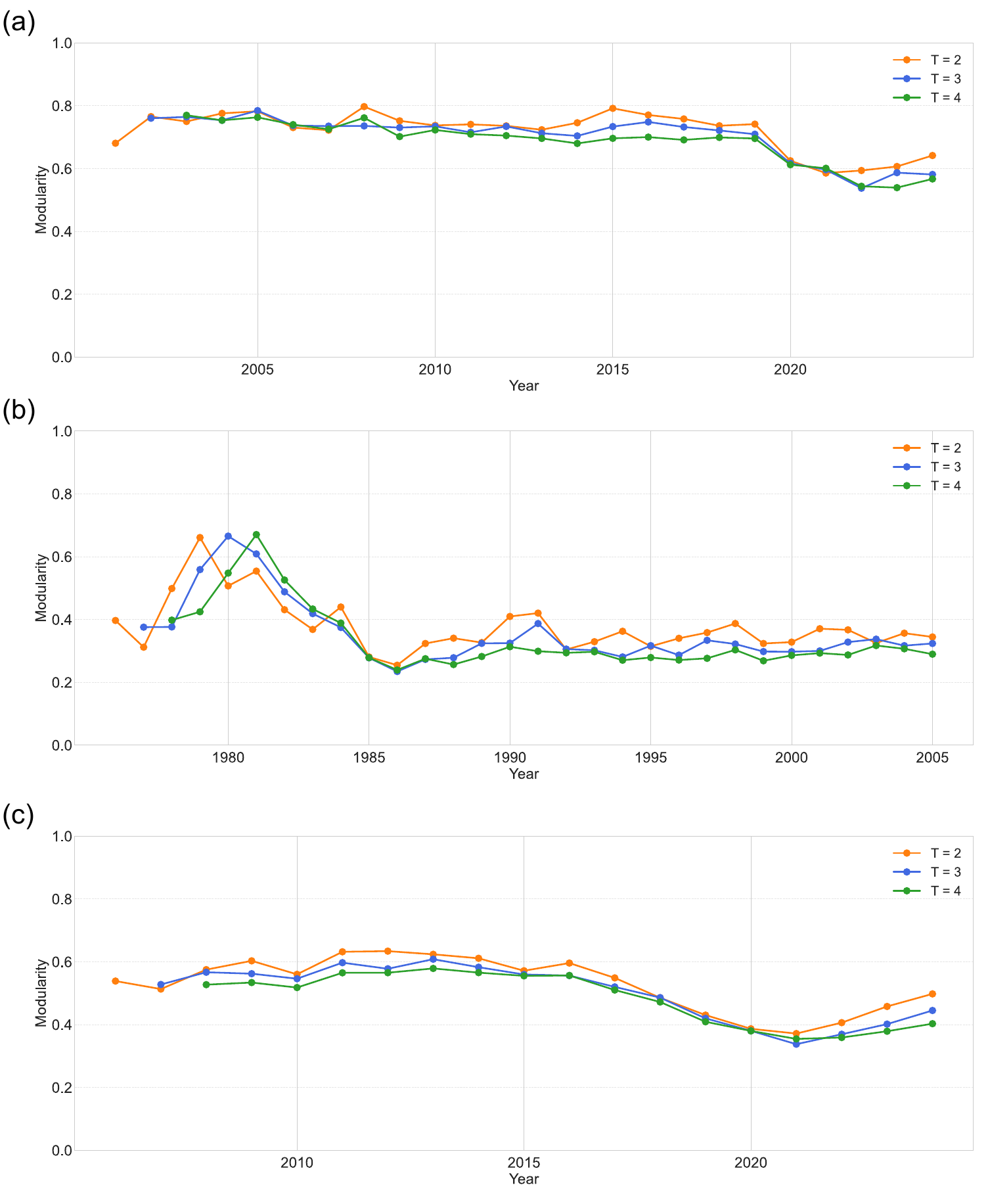}
\caption{Temporal evolution of modularity for time-window widths $T = 2, 3, 4$ years (point estimates). (a) Higher-order network science. (b) Superstring theory. (c) Graph representation learning. Each curve shows the modularity of the observed weighted co-citation network (co-citation frequency $\geq 1$) in each time window (a single estimate without bootstrapping).}
\label{fig:s2}
\end{figure}

The time-series analysis in the main text uses a time-window width of $T = 3$ years.
In this section, we confirm that the marked modularity declines do not depend on this choice.
Using the same weighted co-citation network as in the main text (co-citation frequency $\geq 1$), we computed the modularity of the observed co-citation network in each time window (a single network without bootstrapping) for window widths of $T = 2, 3, 4$ years.
Figure~\ref{fig:s2} shows the temporal evolution of modularity in each research area.

In all research areas, the timing of the marked modularity declines was nearly identical across $T = 2, 3, 4$: the sharp decline in 2020 in higher-order network science, the decline in 1985 in superstring theory, and the sustained decline from 2017 through 2021 in graph representation learning were reproduced for every window width.
The subsequent rise in 2022 and beyond in graph representation learning was also reproduced for every window width.
Because the number of source papers in each window changes with the window width, the absolute level of modularity varies slightly.
However, the timing of the declines and the overall trajectories are robust to the choice of window width.

\section{Robustness of the Bootstrap Variation of Modularity\\ to the Stochasticity of the DCSBM Inference}
\label{sec:seed}

The time-series analysis in the main text interprets the spread of the bootstrap empirical distribution in each time window as the sensitivity of modularity to the resampling of the source-paper set.
However, because the MCMC inference of the DCSBM is itself stochastic, part of this spread may stem from the stochasticity of the inference algorithm.
In this section, we separate the two contributions.

For each research area, we consider the same time window as in Section~\ref{sec:loo} (the time window analyzed for the contribution score $\delta_u$; Table~\ref{table:s_loo}).
Without resampling, we constructed and fixed the weighted co-citation network (co-citation frequency $\geq 1$, largest connected component) from the entire observed source-paper set, varied only the MCMC random seed, and repeated the weighted DCSBM inference (discrete-Poisson weight model with degree correction) $10^3$ times.
Table~\ref{table:s_seed} shows the seed-to-seed standard deviation and 95\% interval of modularity, alongside the standard deviation and 95\% bootstrap interval from the same bootstrap as in the main text ($B = 10^3$).
The seed-to-seed mean modularity of the fixed network differs from the bootstrap mean at the level of a few percent.
This difference is expected, as resampling changes the network itself.

The seed-to-seed standard deviations are $0.005$--$0.008$, which is $3.6$--$10$ times smaller than the bootstrap standard deviations in the same time windows ($0.017$--$0.065$).
In terms of variance, the stochasticity of the MCMC accounts for at most about $8\%$ of the bootstrap variation.
The spread of the bootstrap intervals reported in the main text therefore stems mainly from the resampling of the source papers.
We note that even on the same fixed network, the number of detected communities varies considerably across seeds (for example, from $327$ to $1{,}369$ in the higher-order network science window), whereas the variation of modularity remains small.

\begin{table}[h]
\centering
\small
\caption{Comparison between the seed-to-seed variation of modularity under the MCMC of the DCSBM inference and the bootstrap variation, in the time window analyzed for the contribution score $\delta_u$ in each research area. The seed-to-seed rows show the results of $10^3$ repetitions of the weighted DCSBM inference, in which only the MCMC seed was varied, on the fixed weighted co-citation network (co-citation frequency $\geq 1$, largest connected component; $N$ nodes and $M$ edges) constructed from the entire observed source-paper set. The bootstrap rows show the same results as the main-text time-series analysis ($B = 10^3$). All 95\% intervals are the 2.5th and 97.5th percentiles of the empirical distributions.}
\label{table:s_seed}
\begin{tabular}{lccc}
\toprule
 & \shortstack{Higher-order\\network science} & \shortstack{Superstring\\theory} & \shortstack{Graph representation\\learning} \\
\midrule
Time window & 2018--2020 & 1983--1985 & 2015--2017 \\
Number of nodes & 32{,}532 & 1{,}764 & 15{,}568 \\
Number of edges & 986{,}543 & 42{,}532 & 400{,}272 \\
\midrule
\multicolumn{4}{l}{\textbf{Seed-to-seed (fixed network)}} \\
Mean & 0.620 & 0.289 & 0.525 \\
Standard deviation & 0.006 & 0.008 & 0.005 \\
95\% interval & [0.612, 0.629] & [0.279, 0.299] & [0.515, 0.534] \\
\midrule
\multicolumn{4}{l}{\textbf{Bootstrap ($B=10^3$)}} \\
Mean & 0.658 & 0.330 & 0.556 \\
Standard deviation & 0.065 & 0.050 & 0.017 \\
95\% interval & [0.519, 0.746] & [0.230, 0.420] & [0.525, 0.587] \\
\bottomrule
\end{tabular}
\end{table}

\section{Robustness of the Results to the Choice\\ of the Community-Detection Algorithm}
\label{sec:leiden}

The main analyses detect communities with the DCSBM, which fits a generative model under the minimum description length criterion.
In this section, we examine whether the results depend on this choice by repeating the weighted bootstrap analysis with the Leiden algorithm~\citep{traag2019}, which maximizes modularity directly.
For each research area and each time window, we applied the Leiden algorithm to the same $B = 10^3$ bootstrap samples as in the main analysis and recomputed the modularity distributions, the contribution scores $\delta_u$, and Cliff's $\delta$ between adjacent time windows.

The two algorithms differ substantially in the absolute level of modularity and in the granularity of the partitions: the Leiden algorithm yields higher modularity and fewer communities, whereas the DCSBM yields finer partitions under the description-length control.
The two effect-size measures of the framework, however, compare modularity distributions within a single algorithm: Cliff's $\delta$ compares adjacent time windows, and $\delta_u$ compares the bootstrap samples containing and not containing a given source paper.
The level difference therefore does not enter these measures.

\begin{figure}[p]
\centering
\includegraphics[width=\linewidth]{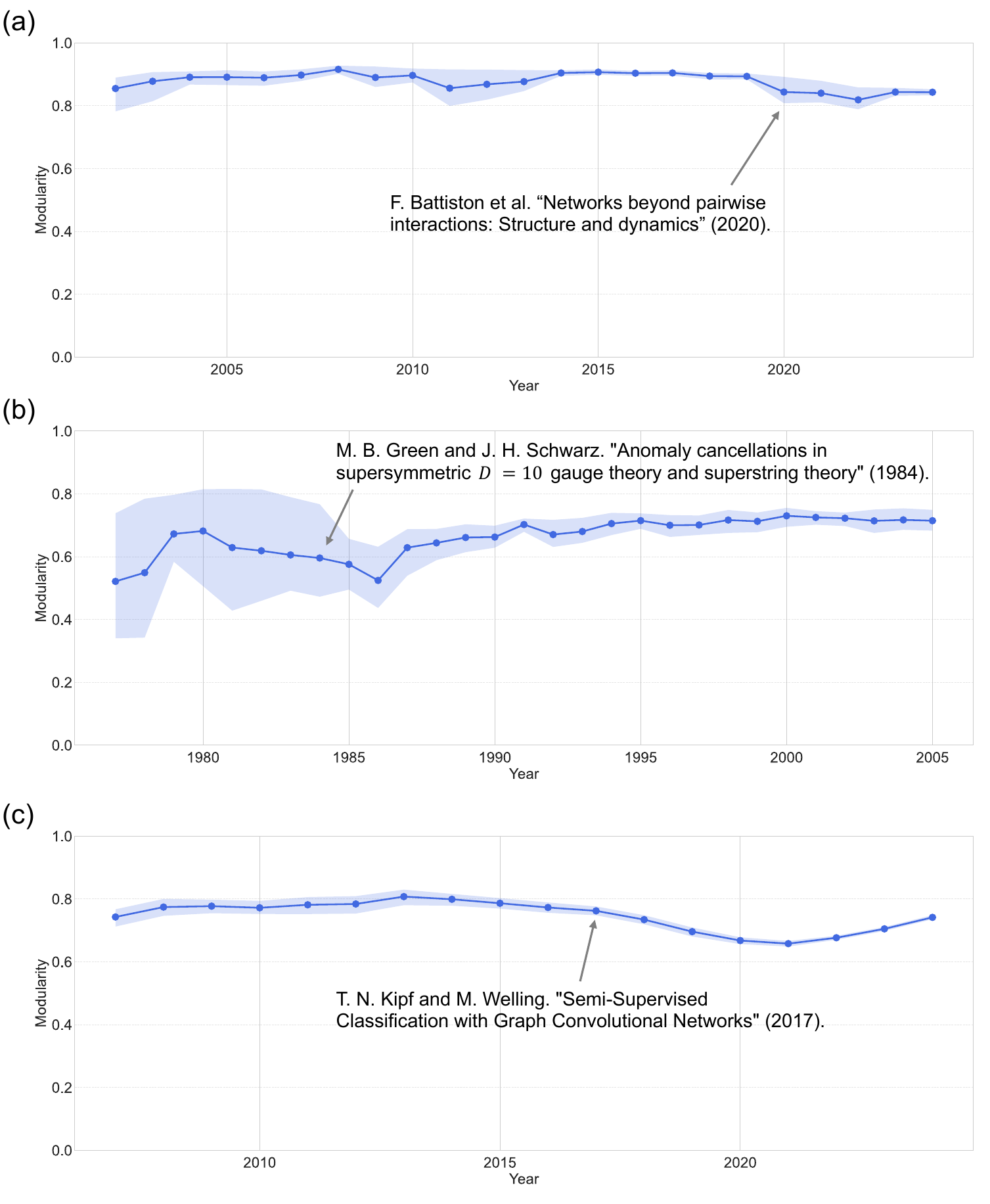}
\caption{Time evolution of the modularity of the intellectual bases under the Leiden algorithm, the counterpart of Fig.~3 in the main text. (a) Higher-order network science. (b) Superstring theory. (c) Graph representation learning. Solid lines show the means over 1,000 bootstrap samples, and shaded areas show the 95\% bootstrap intervals; the bootstrap samples are identical to those in the main analysis. Each three-year time window is plotted at its final year.}
\label{fig:s3}
\end{figure}

Figure~\ref{fig:s3} shows the temporal evolution of modularity computed by the Leiden algorithm, the counterpart of Fig.~3 in the main text.
The declines observed in the main text are largely reproduced: modularity declines in 2020 and beyond in higher-order network science, declines through the mid-1980s with a local minimum in the 1984--1986 window in superstring theory, and declines steadily from the mid-2010s through 2021 in graph representation learning; the subsequent rise in 2022 and beyond is also reproduced.

The contribution scores identify the same key papers.
In every research area, the source paper with the largest $\delta_u$ is identical between the two algorithms: the review by Battiston et al.~\citey{battiston2020} in higher-order network science ($\delta_u > 0.99$ under both algorithms), the lecture notes by J.~Ellis~\citey{ellis1985supersymmetry} in superstring theory ($0.82$ under the DCSBM and $> 0.99$ under the Leiden algorithm), and the review on geometric deep learning by Bronstein et al.~\citey{bronstein2017geometric} in graph representation learning ($0.37$ and $0.26$).
The Spearman rank correlation between the two sets of $\delta_u$ over the source papers in the analyzed window is $0.71$ in higher-order network science, $0.65$ in superstring theory, and $0.59$ in graph representation learning.
The distinction between review-led integration in higher-order network science and collective integration in the other two areas is thus largely preserved under the Leiden algorithm.

For the time series, the Spearman rank correlation of Cliff's $\delta$ across all adjacent-window pairs is $0.75$ in higher-order network science, $0.67$ in superstring theory, and $0.97$ in graph representation learning.
The declines at the main transitions are largely reproduced with large effect sizes under both algorithms: the 2019 vs 2020 and 2021 vs 2022 pairs in higher-order network science ($\delta = 0.88$ and $0.54$ under the DCSBM; $0.93$ and $0.50$ under the Leiden algorithm), the consecutive large effects through 2020 vs 2021 in graph representation learning (from 2016 vs 2017 under the DCSBM and from 2014 vs 2015 under the Leiden algorithm), and the 1991 vs 1992 pair in superstring theory ($0.90$ and $0.66$).
The rise after the 2019--2021 window in graph representation learning is likewise consistent between the algorithms ($\delta = -0.90$, $-0.99$, and $-1.00$ under the DCSBM; $\delta < -0.99$ for all three pairs under the Leiden algorithm).
The two algorithms localize the decline of the mid-1980s in superstring theory differently: the DCSBM shows large effects at 1984 vs 1985 ($0.58$) and 1985 vs 1986 ($0.51$), whereas the Leiden algorithm concentrates the decline at 1985 vs 1986 ($\delta = 0.467$), marginally below the criterion for a large effect size.
Comparisons between the algorithms are therefore better read from the continuous values rather than from the binary classification at $\delta \geq 0.47$.
For the same reason, we do not mark the $\delta \geq 0.47$ classification in Fig.~S3.

The Leiden variant is computationally far cheaper: in our environment, community detection on a single bootstrap sample took a few seconds with the Leiden algorithm and a few minutes with the DCSBM inference, a difference of up to two orders of magnitude.
The Leiden algorithm therefore offers a practical alternative when the framework is applied to larger corpora, while the DCSBM remains the reference method of this study.

\newpage